\documentclass[english,aps,pra,10pt,tightenlines,fleqn,twocolumn,superscriptaddress,floatfix]{revtex4-2}
\renewcommand{\bibliography}[1]{}
\pdfoutput=1
\usepackage[utf8]{inputenc}
\usepackage[T1]{fontenc}
\usepackage{amsmath,amssymb,amsfonts}
\usepackage{bm,bbm}
\usepackage{epsfig}
\usepackage{times}
\usepackage{graphics}
\usepackage{booktabs}

\usepackage[dvipsnames]{xcolor}
\definecolor{dblue}{rgb}{0,0.1,.6}

\usepackage[colorlinks=true,citecolor=dblue,linkcolor=dblue,urlcolor=dblue]{hyperref}
\usepackage[all]{hypcap}

\usepackage{enumitem}
\setlist[enumerate]{label=(\alph*), itemsep=0pt, leftmargin=*}
\newcommand{\ud}{\mathrm{d}}
\newcommand{\id}{\mathbbm{1}}
\newcommand{\Tr}{\operatorname{Tr}}
\newcommand{\bra}{\langle}
\newcommand{\ket}{\rangle}
\newcommand{\mc}[1]{\mathcal{#1}}
\newcommand{\pdag}{{\phantom{\dag}}}
\renewcommand{\vec}[1]{{\boldsymbol{#1}}}
\newcommand{\mri}{\mathrm{i}\mkern1mu}
\DeclareMathOperator*{\argmin}{argmin}

\newcommand{\A}{\mc{A}}
\newcommand{\B}{\mc{B}}
\renewcommand{\O}{\mc{O}}

\newcommand{\hs}{{\hat{\sigma}}}
\newcommand{\dm}{{\hat{\rho}}}
\newcommand{\dml}{{\hat{\varrho}}}
\newcommand{\ha}{\hat{a}}
\newcommand{\ta}{\tilde{a}}
\newcommand{\hn}{\hat{\nu}}
\newcommand{\hN}{\hat{N}}
\newcommand{\hC}{\hat{C}}
\newcommand{\hD}{\hat{D}}
\newcommand{\hF}{\hat{F}}
\newcommand{\hG}{\hat{G}}
\newcommand{\hK}{\hat{K}}
\newcommand{\hP}{\hat{P}}
\newcommand{\hR}{\hat{R}}
\newcommand{\hU}{\hat{U}}
\newcommand{\hw}{\hat{w}}

\newcommand{\vn}{\vec{\nu}}
\newcommand{\vs}{\vec{s}}
\newcommand{\vm}{\vec{m}}
\newcommand{\valpha}{\vec{\alpha}}

\renewcommand{\Re}{\operatorname{Re}}
\renewcommand{\Im}{\operatorname{Im}}
\renewcommand{\Pr}{\operatorname{Pr}}
\newcommand{\veps}{\varepsilon}
\newcommand{\CC}  {\mathbb{C}}
\newcommand{\EE}  {\mathbb{E}}
\newcommand{\SWAP}{\text{SWAP}}
\newcommand{\CNOT}{\text{CNOT}}
\newcommand{\Had} {\text{H}}
\newcommand{\diag}{{\mathrm{diag}}}
\newcommand{\Bra}{\left\langle}
\newcommand{\Kets}{\right\rangle_\text{shots}}
\newcommand{\Nshots}{\mc{N}_\text{s}}
\newcommand{\est} {\text{est}}
\newcommand{\err} {\text{err}}

\newcommand  {\Pmatrix}[1]{\begin{pmatrix}#1\end{pmatrix}}

\makeatletter
\renewcommand{\thesubsection}{\thesection.\Alph{subsection}}
\def\p@subsection{}
\renewcommand{\@seccntformat}[1]{\csname the#1\endcsname\def\@tempa{#1}\def\@tempb{subsection}\ifx\@tempa\@tempb\else.\fi\quad}
\let\oldappendix\appendix
\renewcommand{\appendix}{ \oldappendix \renewcommand{\thesubsection}{\thesection.\arabic{subsection}} }
\makeatother

\newcommand{\qlab}  {National Quantum Laboratory, University of Maryland, College Park, MD 20742, USA}
\newcommand{\umd}   {Department of Physics, University of Maryland, College Park, MD 20742, USA}
\newcommand{\duke}  {Department of Physics and Duke Quantum Center, Duke University, Durham, NC 27708, USA}

\begin{document}

\title{\texorpdfstring{Replica-based holographic measurement schemes for R\'{e}nyi entanglement entropies\\ on classical and quantum computers}{Replica-based holographic measurement schemes for R\'{e}nyi entanglement entropies on classical and quantum computers}}
\author{Thomas Barthel}
\affiliation{\qlab}
\affiliation{\umd}
\affiliation{\duke}
\date{September 3, 2026}

\begin{abstract}
The multiscale entanglement renormalization ansatz (MERA) is believed to efficiently capture low-energy states of strongly correlated quantum many-body systems and can be implemented on quantum computers with a system-size independent number of qubits using tensor Trotterization. This work describes techniques for measuring bipartite R\'{e}nyi entanglement entropies of MERA states on classical computers and quantum computers using the replica trick. The causal structure of MERA enables a \emph{holographic R\'{e}nyi measurement}, where integer-order entropies are determined by the expectation value of cyclic shift operators on renormalized sites at the edge of causal-cone tensor networks for the subsystem boundary. On quantum computers, the measurement can be implemented for any R\'{e}nyi order using fermionic Fourier transforms and parity-controlled flux operators. For the R\'{e}nyi-2 entropy, simple circuits of Hadamard and CNOT gates suffice, and we describe similar custom circuits for the R\'{e}nyi-3 and R\'{e}nyi-4 entropies. We substantiate a time-complexity advantage for the replica-based scheme over state tomography as well as classical shadows with uniform local Pauli measurements.
\end{abstract}

\maketitle

\section{Introduction}\label{sec:intro}
\begin{figure*}[t]
	\includegraphics[width=\textwidth]{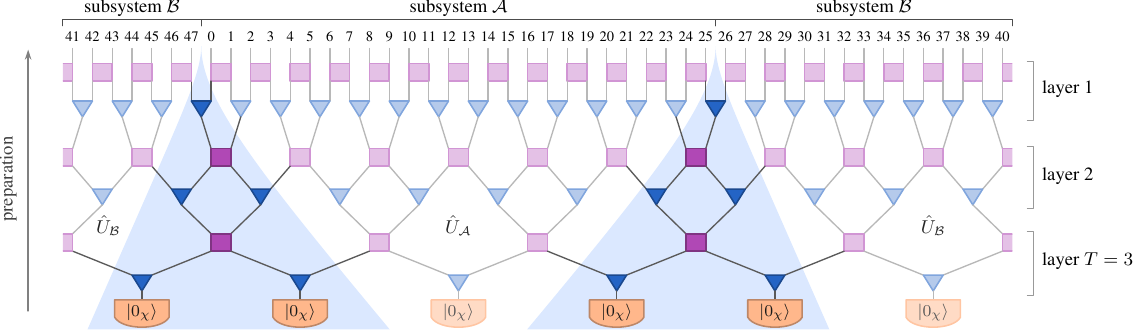}
	\caption{\label{fig:MERA}Binary MERA for a 1D system with 48 sites and periodic boundary conditions that is split into subsystem $\A=\{0,\dotsc,25\}$ and its complement $\B=\{26,\dotsc,47\}$. To access the corresponding order-$n$ R\'enyi entropy \eqref{eq:Renyi}, we only need to implement ($n$ copies of) the unshaded tensors, which constitute the causal cone of the subsystem boundary $\partial\A$. Removing all shaded (isometric) tensors of the MERA outside of the boundary's causal cone corresponds to inverse basis transformations $\hU_\A^\dag$ and $\hU_\B^\dag$ in the two subsystems, which do not affect the subsystem entropy. The preparation direction is upwards. Isometries $\hw:\CC^\chi\to\CC^\chi\otimes\CC^\chi$ (triangles) can be implemented by a unitary $\hat{u}:\CC^\chi\otimes\CC^\chi\to\CC^\chi\otimes\CC^\chi$ and an ancilla in a reference state $|0_\chi\ket$ such that $\hw=\hat{u}\cdot(\id_\chi\otimes|0_\chi\ket)$. In the top layers, $\chi$ reduces until reaching the physical-site Hilbert space dimension in the output of layer 1.}
\end{figure*}
Entanglement entropies characterize quantum many-body states beyond local observables and can probe quantum critical properties. Ground states of local Hamiltonians typically obey entanglement area or log-area laws \cite{Eisert2008,Latorre2009,Laflorencie2016-646}, while universal subleading terms can reveal topological order \cite{Kitaev2006-96,Levin2006-96}. Examples for more exotic behavior are deformed Motzkin and Fredkin chains which display volume-law groundstate entanglement despite having local interactions \cite{Zhang2017-114,Salberger2017-6}. Resolving these scaling laws requires access to entanglement over a relatively large range of subsystem sizes. This is challenging on both classical and quantum computers because entropies depend nonlinearly on the reduced density operator and are not directly accessible through local expectation values.

This work describes holographic classical and quantum algorithms to determine integer-order R\'enyi entanglement entropies in strongly correlated systems using the replica trick. Specifically, let us consider lattice systems in $D$ spatial dimensions with finite-dimensional site Hilbert spaces, including spin systems, fermions in the tight-binding approximation, or corresponding bosonic systems with a local Hilbert space truncation. The multi-scale entanglement renormalization ansatz (MERA), introduced by Guifr\'{e} Vidal for classical simulations \cite{Vidal-2005-12,Vidal2006,Evenbly2009-79}, has two decisive advantages for the task at hand: It is believed to capture low-energy states of strongly correlated systems, and its causal structure will allow us to strongly reduce resource requirements compared with direct preparation and analysis of the physical states of interest.

Tensor networks represent many-body states as networks of partially contracted tensors and provide efficient approximations for many physically relevant low-energy states. A MERA for a $D$-dimensional system is a hierarchical tensor network in $D+1$ dimensions, where the additional dimension corresponds to different layers of tensors and is associated with real-space renormalization steps as in Kadanoff's block spin transformation \cite{Kadanoff1966-2,Jullien1977-38,Drell1977-16}. Every layer decimates the number of (renormalized) sites by a certain coarse-graining ratio \cite{Vidal-2005-12,Vidal2006,Evenbly2009-79} as illustrated in Fig.~\ref{fig:MERA}. Each renormalized site is associated with a local Hilbert space of dimension $\chi$ which, simultaneously, is the MERA's \emph{bond dimension}. The MERA has beneficial properties for both classical and quantum computation:
\begin{enumerate}
 \item One can construct suitable MERA for any number $D$ of spatial dimensions \cite{Vidal-2005-12,Vidal2006,Evenbly2009-79,Cincio2008-100,Evenbly2009-102}, capturing entanglement area laws for all $D$ and log-area laws for $D=1$ \cite{Barthel2010-105,Evenbly2011_06,Evenbly2014-89a}, and one can work directly in the thermodynamic limit with homogeneous MERA,
 \item all MERA tensors are unitary or isometric, which is favorable for an implementation on quantum computers -- e.g., using tensor Trotterization~\cite{Miao2021_08,Miao2023_03,Haghshenas2022-12},
 \item MERA expectation values for local observables and correlators depend only on narrow \emph{causal cones} of tensors \cite{Vidal2006,Evenbly2009-79} such that, in contrast to projected entangled-pair states (PEPS) \cite{Verstraete2004-7,Verstraete2006-96,Niggemann1997-104,Nishino2000-575}, they can be evaluated exactly with polynomial cost on classical computers and one can study large systems on small quantum processors with a system-size independent number of qubits and logarithmically scaling circuit depths \cite{Miao2021_08,Miao2023_03,Haghshenas2022-12,Haghshenas2023_05,Miao2026-17},
 \item local MERA observables are robust to noise~\cite{Kim2017_11},
 \item MERA optimizations are not hampered by barren plateaus \cite{Barthel2023_03,Miao2024-109},
 \item one does not need to handle non-local Jordan-Wigner strings when applying MERA to fermionic systems \cite{Barthel2009-80,Corboz2009-80,Kraus2009_04,Corboz2009_04,Pineda2009_05}, and
 \item the set of MERA is closed, which implies that optimizers always exist \cite{Barthel2022-112}.
\end{enumerate}
The classical computation times for the evaluation of local observables and energy optimization steps for MERA scale as $\O(\chi^r)$ in the bond dimension $\chi$. The relatively large exponents $r$, particularly for $D\geq 2$ dimensions, limit accessible bond dimensions \cite{Evenbly2013,Barthel2025-111} and motivate hybrid quantum-classical MERA schemes, where the classically expensive tensor contractions are implemented cheaply on quantum computers \cite{Miao2021_08,Miao2023_03,Haghshenas2022-12,Haghshenas2023_05}.
\begin{figure*}[t]
	\centering
	\includegraphics[width=0.87\textwidth]{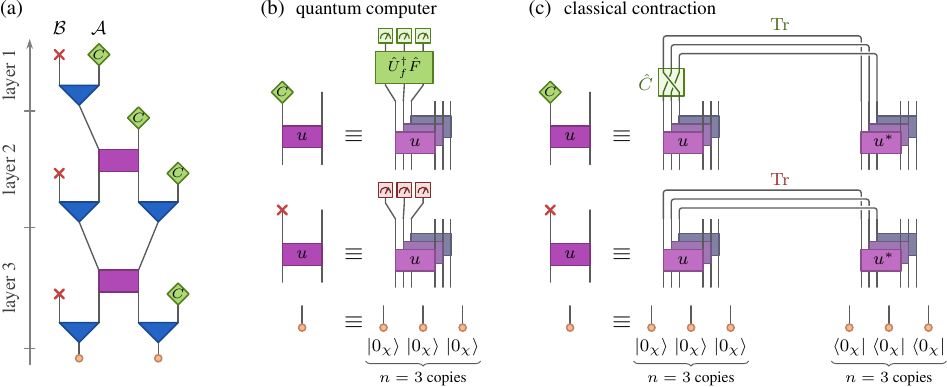}
	\caption{\label{fig:measure}(a) To measure the $n$-th moment $\Tr(\dm_\A^n)=\Tr(\hC_\A\dm_\A^{\otimes n})$, which determines the R\'enyi entropy \eqref{eq:Renyi} using the replica trick, we measure cyclic shift operators $\hC$ on the renormalized sites that leave the causal cone into subsystem $\A$ (green diamonds). Renormalized sites that leave the causal cone into subsystem $\B$ (red crosses) are measured in the computational basis (traced out). This is illustrated here for the left portion of the $\partial\A$ causal cone from Fig.~\ref{fig:MERA}.
	(b) On quantum computers, we prepare $n$ copies of the causal-cone state and $\hC$ can for example be measured using the $\hU_f^\dag \hF$ circuits from Sec.~\ref{sec:shiftTest}.
	(c) In the classical algorithm, based on tensor contractions, we implement $n$ copies of each causal-cone tensor and its complex conjugate.}
\end{figure*}

For a pure state $|\Psi\ket$, the bipartite R\'enyi-$n$ entanglement entropies are
\begin{equation}\label{eq:Renyi}
	S_n(\A) = \frac{1}{1-n}\log \Tr(\dm_\A^n),
    \ \ \text{where}\ \
    \dm_\A = \Tr_\B |\Psi\ket\bra\Psi|
\end{equation}
is the reduced density operator for a subsystem $\A$, $\B$ is the complement of $\A$, and we consider integer orders $n=2,3,\dotsc$. 
Subsystem $\A$ shall be a compact region with volume $\ell^D$ and surface area
\begin{equation}
	|\partial\A|\propto\ell^{D-1}.
\end{equation}
In one dimension, $\A$ is a contiguous interval. In two dimensions, it can, for example, be a square or disk.

Employing the \emph{replica trick} \cite{Edwards1975-5,Holzhey1994-424,Ekert2002-88,Horodecki2002-89,Calabrese2004,Daley2012-109,Islam2015-528}, the R\'{e}nyi entropy \eqref{eq:Renyi} can be determined by preparing $n$ copies of the state $|\Psi\ket$ and measuring the expectation value of the cyclic shift operator $\hC_\A$ among the copies on subsystem $\A$,
\begin{equation}\label{eq:replica}
	\Tr(\dm_\A^n)=\Tr(\hC_\A\dm_\A^{\otimes n}) = \bra\Psi^{\otimes n}|\hC_\A\otimes\id_\B|\Psi^{\otimes n}\ket.
\end{equation}
Experimentally, the replica trick has for example been used in Refs.~\cite{Islam2015-528,Linke2018-98}. Alternatively, nonlinear state properties can be estimated using randomized measurements and classical shadows for individual copies \cite{VanEnk2012-108,Huang2020-16,Brydges2019-364}, and holographic subsystem tomography for MERA has been demonstrated in Ref.~\cite{Miao2026-17}. The time complexity of these alternative approaches depends on both the state representation and the estimator. In terms of general upper bounds and lower bounds for a specific class of MERA, we will substantiate an exponential sample-complexity separation in the number of relevant subsystem-$\A$ qubits between replica measurements and any estimator based on independent, uniformly random Pauli measurements.

A decisive advantage of MERA states for the computation of subsystem entropies is their causal structure: We can remove all tensors outside the causal cone of the subsystem boundary $\partial\A$ \footnote{In the holographic measurement schemes, we can remove all isometries of the MERA outside of the subsystem boundary's causal cone, corresponding to inverse basis transformations $\hU_\A^\dag$ and $\hU_\B^\dag$ in the two subsystems as indicated in Fig.~\ref{fig:MERA}. Importantly, the R\'{e}nyi entanglement entropy \eqref{eq:Renyi} is invariant under such local unitary transformations. The unitary transformations leave us with $K$ nontrivial renormalized qubits in subsystem $\A$, residing at the corresponding edge of the boundary causal cone. The remaining ancillary subsystem-$\A$ qubits are in a product state ($|0,\dotsc,0\ket$) and can be disregarded; in particular, they need not be implemented experimentally \cite{Miao2026-17}.} and only need to measure $\hC_\A$ on the 
\begin{subequations}\label{eq:holoSiteNumber}
\begin{align}
	\O(\log\ell)	& \quad\text{(for one spatial dimension) or}\\
	\O(\ell^{D-1})  & \quad\text{(for $D\geq 2$ spatial dimensions)}
\end{align}
\end{subequations}
renormalized sites at the inner edge of the causal-cone tensor network / quantum circuit as shown in Figs.~\ref{fig:MERA}-\ref{fig:Martini}. Importantly, in classical simulations or on quantum computers with midcircuit measurements and qubit reset \cite{Riste2012-109,Magnard2018-121,Gaebler2021-104}, Eq.~\eqref{eq:holoSiteNumber} does \emph{not} reflect memory requirements, as the renormalized sites can be measured asynchronously and reused.

For 1D and 2D systems, Sec.~\ref{sec:classical} describes the boundary causal-cone tensor networks and their classical contraction costs. Section~\ref{sec:quantum} introduces the quantum measurement schemes, analyzes their time, space, and sample complexities, and compares them with subsystem tomography and classical shadows. Section~\ref{sec:shiftTest} presents cyclic-shift measurement circuits, including custom circuits for $n=2,3,4$ and a general-$n$ construction based on fermionic Fourier transforms and a parity-controlled flux operator. Section~\ref{sec:conclusion} concludes with a discussion. Appendices~\ref{appx:tomog}-\ref{appx:PauliLower} provide derivations for sample complexities, and Appendices~\ref{appx:Uf}-\ref{appx:small-n} give details for the measurement circuits.

The present paper is a companion to the study \cite{Jeanette2026_12}, laying out details of the replica-based holographic R\'{e}nyi measurement schemes. Using a trapped-ion quantum computer and suitable error-mitigation techniques, Ref.~\cite{Jeanette2026_12} presents experimental results for R\'{e}nyi-2 and R\'{e}nyi-3 entanglement entropies in critical and gapped spin chains. Combining MERA with the replica trick gives access to large subsystems and enables the observation of logarithmic growth and area-law saturation, respectively.

\newpage
\section{\texorpdfstring{Tensor-network structure and\\ classical algorithms}{Tensor-network structure and classical algorithms}}\label{sec:classical}
\begin{figure}[t]
	\includegraphics[width=\columnwidth]{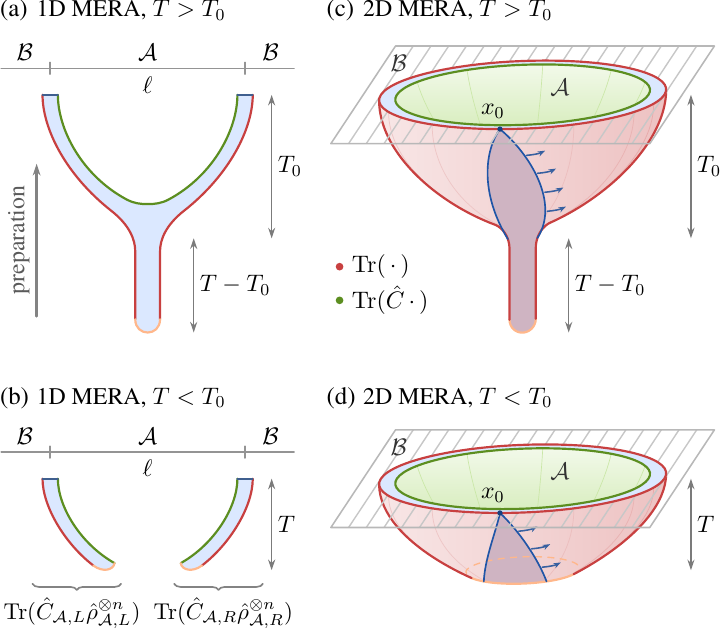}
	\caption{\label{fig:Martini}Structure of MERA tensor networks or tensor network circuits with measurements in classical and quantum simulations, respectively, for the measurement of $\Tr(\dm_\A^n)$ using the replica trick, where the compact subsystem $\A$ has linear size $\ell$. (a) 1D MERA with $T>T_0=\Theta(\log\ell)$ layers and (b) $T<T_0$ as well as (c) 2D MERA with $T>T_0=\Theta(\log\ell)$ and (d) $T<T_0$. Red boundaries of tensor networks indicate that renormalized sites which leave the causal cone (to the outside $\B$) are traced out, i.e., measured in the computational basis. Green ``inner'' boundaries indicate the measurement of cyclic shift operators \eqref{eq:C}.}
\end{figure}

On classical computers, we can stack the $n$ copies of the causal-cone tensor network on top of each other in a $(D+2)$-th dimension and apply $\hC$ as well as trace operations at the edges of the causal cone to obtain the closed tensor network for $\Tr(\dm_\A^n)$ as illustrated in Fig.~\ref{fig:measure}. This closed network then has to be contracted as illustrated in Fig.~\ref{fig:Martini} for 1D and 2D.

In 1D, this tensor network consists of two strands (one for each boundary of the contiguous subsystem $\A$) of height $\min(T,T_0)$ with system-size independent width, where $T$ is the number of MERA layers. If $T$ is larger than a certain minimum $T_0=\Theta(\log\ell)$, the two causal-cone strands will merge into a single strand, continuing for the remaining $T-T_0$ layers. In both cases, the computation time is of order $\Theta(T \chi^{\Theta(n)})$.

Similarly, in 2D with $T\leq T_0=\Theta(\log\ell)$, the obtained closed tensor network has the shape of a ring with circumference $\Theta(\ell)$ and height $T$. Proceeding in the ring direction, it can be contracted with $\Theta(\ell \chi^{\Theta(nT)})$ operations.

In 2D with $T>T_0$, the ring will merge at layer $T_0$ and continue as a single strand for further $T-T_0$ layers such that the complete tensor network has the shape of a Martini glass. The upper part can be contracted as before in the circular fashion with computation time $\Theta(\ell \chi^{\Theta(n\log\ell)})$, and one then contracts the remaining strand with $\Theta\big((T-T_0)\chi^{\Theta(n)}\big)$ operations.

This leads to the classical computation time and space complexities shown in lines 1 and 2 of Tables~\ref{tab:costs-cnct} and \ref{tab:costs-deep}.

\section{Quantum algorithms}\label{sec:quantum}
On quantum computers, the $\chi$-dimensional Hilbert space of each renormalized site of the MERA is implemented using $\lceil\log_2 \chi\rceil$ qubits. Additionally, the MERA tensors are  \emph{Trotterized} \cite{Miao2021_08,Miao2023_03,Haghshenas2022-12}, i.e., each (unitary or isometric) tensor is implemented by a brickwork circuit with $\Theta(t)$ layers, which we refer to as Trotter steps. The reason for the term Trotterization is that the gates of the brickwork circuit get closer to the identity as we increase $t$ and, hence, the precision of the ansatz. The relative scaling of $t$, $\chi$, and groundstate approximation accuracies as well as a resulting quantum advantage are analyzed in Refs.~\cite{Miao2021_08,Miao2023_03,Barthel2025-111}. Alternative Trotterizations are possible, e.g., using parallel random-pair circuits \cite{Miao2023_03}.

All quantum computation time complexities stated in the following refer to parallel circuit depths, not total computational work (number of gates). As an example, we report time complexity $\Theta(n)$ for a brickwork circuit with $\Theta(n)$ layers. The stated space complexities for the holographic schemes assume the availability of midcircuit measurements and qubit reset \cite{Riste2012-109,Magnard2018-121,Gaebler2021-104}, allowing qubits that leave the causal cone to be measured and reused.

\subsection{Full subsystem-state preparation}
The expensive classical tensor contractions now simply correspond to time evolution (gate applications) on the quantum computer. For a MERA with $T$ layers and $t$ Trotter steps, the preparation time is $\Theta(Tt)$. The number $\Nshots$ of required samples is discussed in Sec.~\ref{sec:samples}. If we prepare $n$ copies of the state for subsystem $\A$ fully, we need $\Theta(n\ell^D)$ qubits, where $D$ is the number of spatial dimensions. This full-preparation approach leads to the quantum computation time and space complexities shown in lines 3 and 4 of Tables~\ref{tab:costs-cnct} and \ref{tab:costs-deep}. The additional computation time $\Theta(n)$ is incurred by the measurement circuit for the cyclic shift operator
\begin{equation}\label{eq:Cprod}
	\hC_\A=\hC^{\otimes K}
\end{equation}
It is a tensor product with respect to the $K$ relevant subsystem-$\A$ qubits. For the full subsystem-$\A$ preparation,
\begin{equation}\label{eq:K-full}
	K=K_\text{full}=\Theta(\ell^D).
\end{equation}
The measurement of $\hC_\A$ can be accomplished  using parity-controlled flux operators and fermionic Fourier transforms as detailed in Sec.~\ref{sec:shiftTest}. For $n=2$ R\'{e}nyi entropies, alternative short circuits of Hadamard and CNOT gates suffice.

\subsection{Holographic R\'{e}nyi measurement for 1D}\label{sec:HRM-1D}
As more efficient alternatives to the full subsystem-state preparation we suggest replica-based \emph{holographic R\'{e}nyi measurement} schemes in Secs.~\ref{sec:HRM-1D}-\ref{sec:HRM-2D-B}, where the number $K$ of $\hC$-measurements is strongly reduced to $K_\text{holog}$, scaling according to Eq.~\eqref{eq:holoSiteNumber} times $\log_2\chi$  (the number of qubits per renormalized site). With the number of layers $T_0=\Theta(\log\ell)$ to the branching point,
\begin{equation}\label{eq:K-holog}
	K_\text{holog}=
	\begin{cases}
	\Theta(\min\{T,T_0\}\log\chi)	& \text{for $D=1$},\\
	\Theta(\ell^{D-1}\log\chi)  & \text{for $D\geq 2$}.
	\end{cases}
\end{equation}

Starting the MERA preparation and $\hC_\A$ measurement from (the bottom) layer $T$, we proceed going up, applying the Trotterized tensors of the causal cone for the subsystem boundary $\partial\A$ just as discussed for local observables in Refs.~\cite{Miao2021_08,Miao2023_03}. The only difference is that we apply tensors for $n$ copies of the MERA. Renormalized sites that leave the causal cone are measured (in the computational basis) and the corresponding qubits are reused for the implementation of subsequent isometric tensors. In this way, we proceed until reaching the branching point at layer $T_0=\Theta(\log\ell)$ as shown in Fig.~\ref{fig:Martini}a. This part of the computation requires $\Theta(n\log\chi)$ qubits and $\Theta\big((T-T_0) t\big)$ time (line 5a in Table~\ref{tab:costs-cnct}).

For 1D systems, the circuit then branches into two parts, one for each of the two subsystem boundary points. We proceed as before, except that we now measure cyclic shift operators $\hC$ on the renormalized sites that leave the causal cone into subsystem $\A$ \cite{Note1} as indicated by the green boundaries in Fig.~\ref{fig:Martini}a. The process terminates when reaching the physical layer 1 (top). This part of the computation requires $\Theta(n\log\chi)$ qubits and $\Theta\big(\log\ell\cdot\{t,n\}\big)\equiv \Theta(t \log\ell)+\Theta(n \log\ell)$ time (line 5b in Table~\ref{tab:costs-cnct}).

If the number of MERA layers is $T<T_0$, the circuits for the two boundary points of the subsystem remain independent as shown in Fig.~\ref{fig:Martini}b. In this case, we can use the tensor product structure of the corresponding subsystem state and shift operator to write $\Tr(\dm_\A^n)$ as a product of two terms for the left and right boundary points,
\begin{equation}
	\Tr(\hC_\A\dm_\A^{\otimes n})=\Tr(\hC_{\A,L}\dm_{\A,L}^{\otimes n})\Tr(\hC_{\A,R}\dm_{\A,R}^{\otimes n}).
\end{equation}
Each term can be measured independently, halving the number of required qubits and reducing the computation time to $\Theta\big(T\cdot\{t,n\}\big)$ (line 5b in Table~\ref{tab:costs-deep}).

\subsection{Holographic R\'{e}nyi measurement for 2D -- Scheme~A}\label{sec:HRM-2D-A}
If $T>T_0$, the quantum circuit for 2D systems also starts with a single strand for layers $T$ to $T_0$, which we can implement as in the 1D case with the same time and space complexities (line 6a of Table~\ref{tab:costs-cnct}).

When reaching the branching point at layer $T_0=\Theta(\log\ell)$, the causal-cone circuit of the subsystem boundary $\partial\A$ widens and the total circuit has the form of a Martini glass as shown in Fig.~\ref{fig:Martini}c with bowl height $T_0$ and stem height $T-T_0$.

Following closely the 1D scheme of Sec.~\ref{sec:HRM-1D}, we can continue the circuit implementation from bottom to top. While the time complexity $\Theta\big(\log\ell\cdot\{t,n\}\big)$ is unchanged, we need a larger number of qubits. Specifically, the maximum qubit count of $\Theta(n\ell\log\chi)$ is required when reaching the physical layer (line 6b of Table~\ref{tab:costs-cnct}). Compared to the 1D case, the additional factor $\ell$ results from the fact that the subsystem boundary does not consist of two points anymore but a closed curve of length $\Theta(\ell)$.

If $T<T_0$, the bottom part of the circuit is removed and the circuit has the form of a ring as shown in Fig.~\ref{fig:Martini}d with computational complexity as given in line 6b of Table~\ref{tab:costs-deep}.

\subsection{Holographic R\'{e}nyi measurement for 2D -- Scheme~B}\label{sec:HRM-2D-B}
\begin{table}[t]
	\setlength{\tabcolsep}{1ex}
	\renewcommand{\arraystretch}{1.3}
	\begin{tabular}{r  l  l l}
	\toprule\addlinespace
	Line &	Scheme & Time 		& Space\\[0.2em]
	\midrule\addlinespace
	1)	& 1D classical	& $T \chi^{\Theta(n)}$	& $\chi^{\Theta(n)}$\\[0.3em]
	2a)	& 2D classical	& $(T-T_0)\chi^{\Theta(n)}$	& $\chi^{\Theta(n)}$\\
	2b)	& 				& $\ell \chi^{\Theta(n\log\ell)}$	& $\chi^{\Theta(n\log\ell)}$\\[0.5em]
	\midrule\addlinespace
	3)	& 1D quantum, full prep.	& $\Nshots\cdot\{T t, n\}$		& $n\ell$\\[0.3em]
	4)	& 2D quantum, full prep.	& $\Nshots\cdot\{T t, n\}$		& $n\ell^2$\\[0.5em]
 	\midrule\addlinespace
	5a)	& 1D quantum, holog.	& $\Nshots\cdot(T-T_0) t$		& $n\log\chi$\\
	5b)	& 					& $\Nshots\log\ell\cdot \{t,n\}$		& $n\log\chi$\\[0.3em]
	6a)	& 2D quantum, holog.\ A	& $\Nshots\cdot(T-T_0) t$	& $n\log\chi$\\
	6b)	& 						& $\Nshots\log\ell\cdot \{t,n\}$	& $n\ell\log\chi$\\[0.3em]
	7a)	& 2D quantum, holog.\ B	& $\Nshots\cdot(T-T_0) t$	& $n\log\chi$\\
	7b)	& 						& $\Nshots\log\ell\cdot \{t,n\}$	& $n\log\ell\log\chi$\\
	7c)	& 						& $\Nshots\ell\cdot\{t,n\}$			& $n\log\ell\log\chi$\\[0.3em]
	\bottomrule
	\end{tabular}
	\caption{\label{tab:costs-cnct}Scaling of computation time and space complexities for the measurement of bipartite order-$n$ R\'{e}nyi entanglement entropies on classical and quantum computers with subsystem size $\propto\ell^D$ (surface area $\propto\ell^{D-1}$), MERA bond dimension $\chi$, number of Trotter steps $t$ for tensor Trotterization \cite{Miao2021_08,Miao2023_03}, and the number of MERA layers $T>T_0=\Theta(\log\ell)$, corresponding to panels (a) and (c) of Fig.~\ref{fig:Martini}. $\Nshots$ is the number of required measurement shots given in Eq.~\eqref{eq:Nshots}. The notation $\Theta(a\cdot\{b,c\})$ for a cost scaling means $\Theta(ab)+\Theta(ac)$. See Table~\ref{tab:costs-deep} for the case $T<T_0$.}
\end{table}
\begin{table}[t]
	\setlength{\tabcolsep}{1.1ex}
	\renewcommand{\arraystretch}{1.3}
	\begin{tabular}{r  l  l l}
	\toprule\addlinespace
	Line &	Scheme & Time 		& Space\\[0.2em]
	\midrule\addlinespace
	1)	& 1D classical	& $T \chi^{\Theta(n)}$		& $\chi^{\Theta(n)}$\\[0.3em]
	2b)	& 2D classical	& $\ell \chi^{\Theta(n T)}$	& $\chi^{\Theta(n T)}$\\[0.5em]
	\midrule\addlinespace
	3)	& 1D quantum, full preparation	& $\Nshots\cdot\{T t,n\}$		& $n\ell$\\[0.3em]
	4)	& 2D quantum, full preparation	& $\Nshots\cdot\{T t,n\}$		& $n\ell^2$\\[0.5em]
 	\midrule\addlinespace
	5b)	& 1D quantum, holographic	& $\Nshots T\cdot \{t,n\}$		& $n\log\chi$\\[0.3em]
	6b)	& 2D quantum, holographic\ A	& $\Nshots T\cdot \{t,n\}$	& $n\ell\log\chi$\\[0.3em]
	7b)	& 2D quantum, holographic\ B	& $\Nshots T\cdot \{t,n\}$	& $nT\log\chi$\\
	7c)	& 							& $\Nshots \ell\cdot\{t,n\}$	& $nT\log\chi$\\[0.3em]
	\bottomrule
	\end{tabular}
	\caption{\label{tab:costs-deep}Scaling of computation time and space complexities for order-$n$ R\'{e}nyi measurement in MERA states with subsystem size $\propto\ell^D$ and $T<T_0$ MERA layers, corresponding to panels (b) and (d) of Fig.~\ref{fig:Martini}.}
\end{table}
A downside of Scheme~A for 2D is the increased number of qubits $\Theta(n\ell\log\chi)$. In the more space-efficient Scheme~B, assuming $T>T_0$, we reduce it to $\Theta(n\log\ell\log\chi)$ at the cost of increasing the computation time to $\Theta(\ell\cdot\{t,n\})$ for the bowl of the Martini glass. If $T<T_0$, the qubit count reduces to $\Theta(nT\log\chi)$.

First, assume $T>T_0$. The stem of the Martini-glass shaped circuit is implemented as in Scheme~A (line 7a in Table~\ref{tab:costs-cnct}). Instead of then progressing up beyond layer $T_0$, implementing all tensors for each layer before proceeding to the next layer, we select a random point $x_0$ on the subsystem boundary $\partial\A$ and only implement the tensors and measurements for the causal cone of $x_0$ as shown in Fig.~\ref{fig:Martini}c. This leaves us with $\Theta(n\log\ell\log\chi)$ qubits corresponding to the two edges of the $x_0$ causal-cone circuit as indicated by the blue lines in the figure (line 7b in Table~\ref{tab:costs-cnct}). Then, we proceed from $x_0$ in a circular fashion around the subsystem boundary $\partial\A$, implementing further tensors and measurements until coming back to $x_0$, where the $\hC_\A$ measurement completes. This step requires a computation time $\Theta(\ell\cdot\{t,n\})$ (line 7c in Table~\ref{tab:costs-cnct}).

If $T<T_0$, the stem section of the circuit is removed. As the height of the remaining bowl part is now reduced to $T$, we have to replace $T_0=\Theta(\log\ell)$ by $T$ in the computational complexities (lines 7b and 7c in Table~\ref{tab:costs-deep}).

\subsection{Sample complexity}\label{sec:samples}
Outcomes of quantum measurements are stochastic. To determine the R\'{e}nyi-$n$ entropy \eqref{eq:Renyi} to precision $\veps$, we want an entropy estimator $S_n^\est$ such that
\begin{equation}\label{eq:Sn-precision}
	\Delta S_n:=|S^\est_n-S_n|\leq\veps
\end{equation}
with probability at least $1-p_\err$, where $0<p_\err<\frac{1}{2}$ is a prescribed failure probability that is independent of $n$ and $\ell$.

Let $C_n^\est$ be a real-valued estimator of the $n$-th moment 
\begin{equation}\label{eq:Cn-to-Sn}
	C_n:=\Tr(\dm_\A^n)=\Tr(\hC_\A\dm_\A^{\otimes n})=e^{-(n-1)S_n}.
\end{equation}
For the $K$ relevant subsystem-$\A$ qubits \cite{Note1}, $2^{-K(n-1)}\leq C_n\leq 1$ such that we can define
\begin{subequations}\label{eq:Cn-clipped}
\begin{align}
	C_n^\text{clip}&:=\min\{1,\max\{2^{-K(n-1)},C_n^\est\}\}\quad\text{and}\\
	S^\est_n&:=\frac{1}{1-n}\log C_n^\text{clip}.
\end{align}
\end{subequations}
Clipping cannot increase the moment error, $|C_n^\text{clip}-C_n|\leq|C_n^\est-C_n|$.
According to the inequality $|\log(1+x)|\leq2|x|$ for $|x|\leq\frac{1}{2}$, entropy accuracy \eqref{eq:Sn-precision} with $0<(n-1)\veps\leq 1$ is achieved if
\begin{equation}\label{eq:Cn-precision}
	|C_n^\est-C_n|\leq\delta_n,\ \ \text{where}\quad
	\delta_n:=\frac{1}{2}\,(n-1)\veps C_n.
\end{equation}
Indeed, this condition implies $|C_n^\text{clip}/C_n-1|\leq\frac{1}{2}$ and
\begin{equation}\label{eq:moment-to-entropy}
	\Delta S_n \leq\frac{2|C_n^\text{clip}-C_n|}{(n-1)C_n}
	           \leq\frac{2|C_n^\est-C_n|}{(n-1)C_n}\leq\veps.
\end{equation}

In the replica-based measurement schemes, each shot yields an eigenvalue $\omega$ of the cyclic shift operator $\hC_\A$. As $\hC_\A^n=\id$, these eigenvalues are roots of unity [Eq.~\eqref{eq:Cspectrum} below]. The sample average $C_n^\est:=\Bra\Re\omega\Kets$ is an unbiased real estimator for $C_n$. Due to $|\omega|=1$, $\EE(\omega)=\EE(\Re\omega)=C_n$, and $\EE(\Im\omega)=0$, we have
\begin{equation}
	\EE\left(|\omega-C_n|^2\right) = 1-C_n^2 = \operatorname{Var}(\Re\omega) + \EE\big((\Im\omega)^2\big)
\end{equation}
such that $\operatorname{Var}(\Re\omega)\leq 1-C_n^2$, and the estimator's mean squared error for $\Nshots$ shots is bounded by
\begin{equation}
	\EE\left(|C_n^\est-C_n|^2\right) \leq \frac{1-C_n^2}{\Nshots}\leq\frac{1}{\Nshots}.
\end{equation}
So, using Chebyshev's inequality we have $\Pr(|C_n^\est-C_n|\geq\delta_n)\leq 1/(\Nshots\delta_n^2)$, i.e., we reach the entropy accuracy \eqref{eq:Sn-precision} with probability $\geq 1-p_\err$ when choosing
\begin{equation}\label{eq:Nshots}
	\Nshots
	\stackrel{\eqref{eq:Cn-precision}}{\geq} \frac{4}{p_\err\,(n-1)^2\veps^2C_n^2}
	\stackrel{\eqref{eq:Cn-to-Sn}}{=} \O\left(\frac{e^{2(n-1)S_n}}{(n-1)^2\veps^2}\right).
\end{equation}
As $C_n$ decreases with increasing entanglement, the $\Nshots$ bound \eqref{eq:Nshots} increases accordingly.

For 1D area-law systems, including ground states of gapped local Hamiltonians \cite{Hastings2007-08,Brandao2013-9}, $S_n$ remains bounded when increasing the subsystem size $\ell$, and the $\Nshots$ bound \eqref{eq:Nshots} becomes independent of $\ell$.

In ground states of critical 1D systems governed by a conformal field theory \cite{Francesco1997} with central charge $c$, we have $S_n = \frac{c}{6} \left(1 + \frac{1}{n}\right) \log (\ell/a) + \O(1)$ for a subsystem of length $\ell$ with two entanglement boundaries \cite{Holzhey1994-424,Vidal2003-7,Calabrese2004,Zhou2005-12} such that the sufficient budget $\Nshots \propto \ell^{\frac{c}{3} \left(n - \frac{1}{n}\right)}$ grows polynomially in $\ell$.

For area-law states in $D\geq 2$ spatial dimensions \cite{Eisert2008,Latorre2009,Laflorencie2016-646}, the $\Nshots$ bound is exponential in $\ell^{D-1}$, suggesting that the measurement of subsystem entropies is then generally restricted to subsystems of small linear size $\ell$.

\subsection{Comparison to subsystem tomography and classical shadows}\label{sec:samples-tomog}
Alternatives to using the replica trick and measuring cyclic shift operators on subsystem $\A$ include full quantum state tomography \cite{Vogel1989-40,Hradil1997-55,James2001-64,Haeffner2005-438,BlumeKohout2010-12,Yu2020_09} and randomized moment estimation using classical shadows \cite{VanEnk2012-108,Huang2020-16,Struchalin2021-2,Zhu2022-13,Garratt2024-5} on a single copy. While this reduces the qubit count by a factor $1/n$, tomography and randomized measurements generally determine more information than needed for the measurement of a R\'{e}nyi-$n$ entanglement entropy, which can also lead to an increased sample complexity. Here, we derive general upper bounds on the required sample count $\Nshots$ as well as lower bounds for a specific family of MERA, all displaying an exponential separation from Eq.~\eqref{eq:Nshots} in terms of the number $K$ of relevant subsystem-$\A$ qubits \cite{Note1}. Compared to Sec.~\ref{sec:samples}, different moment estimators $C_n^\est$ are employed, but we use the same conversion \eqref{eq:Cn-clipped} to entropies and assume $0<(n-1)\veps\leq 1$ as well as $0<p_\err<\frac{1}{2}$.

A common practical method for tomography and classical shadows are Pauli measurements on the $K$ subsystem qubits, where axes $\alpha_{s,k}\in\{x,y,z\}$ are chosen independently and uniformly at random for every sample $s$ and qubit $k$. The measurement outcomes $m_{s,k}=\pm 1$ define the classical snapshots
\begin{equation}\label{eq:PauliSnapshot}
	\dm_s:=\bigotimes_{k=1}^K \frac{1}{2}\left(\id+3m_{s,k}\hs^{\alpha_{s,k}}\right)
	\ \ \text{such that}\ \
	\EE(\dm_s)=\dm_\A
\end{equation}
with Pauli operators $\{\hs^x,\hs^y,\hs^z\}$. The factor $3$ compensates for the random axis choice and the snapshot operators $\dm_s$ need not be positive.

For full state tomography, we can project the sample average $\bar{\rho}:=\Bra\dm_s\Kets$ onto the convex set of density operators by minimizing the Frobenius-norm distance to obtain a physical estimator $\dm_\est$ for $\dm_\A$. The estimator $\dm_\est$ can then be used to evaluate subsystem observables, entanglement entropies, the entanglement spectrum, etc. As detailed in Appendix~\ref{appx:tomog}, trace-norm accuracy $\|\dm_\est-\dm_\A\|_1\leq\eta$ can be achieved with $\Nshots^\est(\eta) = \O\left(10^K/\eta^2\right)$, where Markov's inequality is used, and the exponential factor follows from $\|\Delta\dm\|_1\leq\sqrt{2^K}\|\Delta\dm\|_2$ and $\Tr(\dm_s^2)=5^K$.
For the moment estimator
\begin{equation}\label{eq:Cn-tomog}
	C_n^\est:=\Tr(\dm_\est^n)
\end{equation}
the telescoping sum $\dm_\est^n-\dm_\A^n=\sum_{m=0}^{n-1}\dm_\est^{n-1-m}(\dm_\est - \dm_\A)\dm_\A^m$ gives $|C_n^\est-C_n|\leq n\|\dm_\est-\dm_\A\|_1$. Thus, choosing reconstruction accuracy $\eta=\delta_n/n$ and applying Eq.~\eqref{eq:Cn-precision}, one finds that the number of samples needed to achieve $\Delta S_n\leq \veps$ is bounded by
\begin{equation}\label{eq:Nshots-tomog}
	\Nshots^\text{tomo} = \O\left( \frac{10^K n^2 e^{2(n-1)S_n}}{(n-1)^2\veps^2} \right).
\end{equation}

Classical shadows allow for a direct estimation of density operator moments \eqref{eq:Cn-to-Sn} from the measurement records, avoiding the reconstruction of an accurate density operator. An unbiased estimator is
\begin{equation}\label{eq:Cn-shadow}
	C^\est_n :=\frac{(\Nshots-n)!}{\Nshots!}\sum_{\vs}\Re\Tr(\dm_{s_1}\dots\dm_{s_n}),
\end{equation}
where the sum runs over all $\Nshots!/(\Nshots-n)!$ ordered tuples $\vs=(s_1,\dotsc,s_n)$ of distinct sample indices $s_j\in\{1,\dotsc,\Nshots\}$ and $\EE(C^\est_n)=C_n$.
As shown in Appendix~\ref{appx:shadow}, the deviations $d_\vs:=\Re\Tr(\dm_{s_1}\dots\dm_{s_n})-C_n$ for tuples $\vs$ in Eq.~\eqref{eq:Cn-shadow} obey $|\EE(d_\vs d_{\vs'})|\leq 4^{rK}$ if $\vs$ and $\vs'$ share $r$ sample indices. In conjunction with the bound $p_r \leq \binom{n}{r}\left({2n}/{\Nshots}\right)^r$ on the fraction of such tuple pairs, one finds $\EE|C^\est_n-C_n|^2 \leq {4n^2 4^K}/{\Nshots}$ for the mean squared deviation of the estimator \eqref{eq:Cn-shadow} if $\Nshots\geq 2n^2 4^K$. Markov's inequality with $\delta_n$ from Eq.~\eqref{eq:Cn-precision} then gives the sufficient sample count
\begin{equation}\label{eq:Nshots-shadow}
	\Nshots^\text{shadow} = \O\left( \frac{4^K n^2 e^{2(n-1)S_n}}{(n-1)^2\veps^2} \right).
\end{equation}

Both approaches benefit from the holographic reduction due to the causal structure of MERA \cite{Miao2026-17}. Replacing the physical subsystem $\A$ by the renormalized sites at the inner edge of its boundary causal cone preserves its nonzero density-matrix eigenvalues and reduces $K$ from $K_\text{full}$ to $K_\text{holog}$ as stated in Eq.~\eqref{eq:K-holog} \cite{Note1}. Both bounds~\eqref{eq:Nshots-tomog} and \eqref{eq:Nshots-shadow} are \emph{sufficient} upper bounds, but do not assert optimality. Nevertheless, the factors $10^K$ and $4^K$ suggest an exponential overhead in terms of the number $K$ of relevant subsystem-$\A$ qubits compared to the scaling \eqref{eq:Nshots} for the replica-based R\'{e}nyi measurement.

A \emph{necessary} exponential overhead can, for example, be established for a family of ternary MERA with $T$ layers, bond dimension $\chi=2$, and identity disentanglers, i.e., ternary tree tensor networks \cite{Shi2006-74,Tagliacozzo2009-80,Barthel2026-9}: Subsystem $\A$ shall consist of the first $\ell=(3^T+1)/2$ sites. As detailed in Appendix~\ref{appx:PauliLower}, the MERA is chosen to prepare the holographic subsystem state
\begin{equation}\label{eq:PauliLower-state}
\begin{split}
	\dm_{\A,\beta}=\frac{1}{2}\Big(&|0^K\ket\bra0^K|+|1^K\ket\bra1^K|\\
	&+\beta|0^K\ket\bra1^K|+\beta|1^K\ket\bra0^K|\Big),
\end{split}
\end{equation}
where $|\nu^K\ket:=|\nu\ket^{\otimes K}$, $K=T+1$, and the MERA depends on the parameter $\beta\in[1/3,2/3]$ which is unknown to the entropy estimator. The two nonzero $\dm_\A$-eigenvalues $\lambda_\pm=\frac{1}{2}(1\pm\beta)$ result in $C_n=\lambda_+^n+\lambda_-^n$ such that $S_n\leq\log 2$ for all $\beta$ and $T$.
The Pauli measurement distribution depends on $\beta$ only if all axes are $x$ or $y$ and their number $N_y$ of $y$ axes is even. These informative settings occur with probability $q=2^{K-1}/3^K$. For informative settings, the signed parity $\Pi:=(-1)^{N_y/2}\prod_k m_k$ has probabilities $\Pr(\Pi=\pm1)=\frac{1}{2}(1\pm\beta)$. The binary hypothesis testing argument in Appendix~\ref{appx:PauliLower} shows that, for any fixed $n \geq 2$ and $0<p_\err<\frac{1}{2}$, achieving entropy accuracy $\veps$ with probability at least $1-p_\err$ requires
\begin{equation}\label{eq:Nshots-PauliLower}
	\Nshots^\text{Pauli} = \Omega\left(\frac{(3/2)^K}{\veps^2}\right)
\end{equation}
shots for sufficiently small $\veps$. This lower bound holds for any classical processing of the Pauli measurement records, including the state-tomography and classical-shadow estimators \eqref{eq:Cn-tomog} and \eqref{eq:Cn-shadow}. In comparison, Eq.~\eqref{eq:Nshots} with $C_n\geq 2^{1-n}$ gives $\Nshots^\text{replica}=\O(1/\veps^2)$, which establishes an exponential time-complexity separation in terms of $K=\Theta(\log\ell)$.

\section{Shift-operator measurement}\label{sec:shiftTest}
For the replica-based R\'{e}nyi measurement on quantum computers, we finally need to determine circuits allowing us to efficiently measure the cyclic shift operator $\hC_\A=\hC^{\otimes K}$ among the $n$ copies on subsystem $\A$.

\subsection{Schemes for small-\texorpdfstring{$n$}{n} entropies}\label{sec:shiftTest-small-n}
\begin{figure}[t]
	\centering
	\includegraphics[width=0.97\columnwidth]{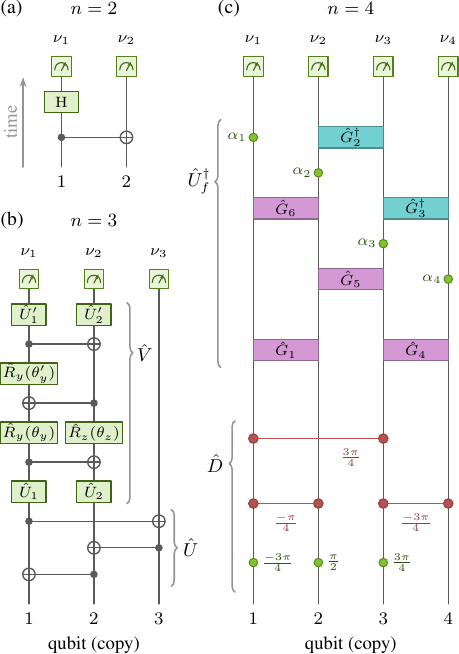}
	\caption{\label{fig:shiftTest-small-n}Custom circuits for measuring the cyclic shift operator $\hC$ from Eqs.~\eqref{eq:Cprod} and \eqref{eq:C} for (a) $n=2$, (b) $n=3$, and (c) $n=4$ copies. Time progresses upwards, and all qubits are measured in the computational basis after applying the shown circuits that diagonalize $\hC$. (a) The destructive SWAP test applies $\Had_1\CNOT_{1,2}$. (b) The $n=3$ circuit applies $(\hat{V}\otimes\id)\hU$, using six CNOT gates. The comprised single-qubit gates and rotation angles are specified in Appendix~\ref{appx:small-3}. (c) The $n=4$ circuit applies $\hU_f^\dag\hD$ from Appendix~\ref{appx:small-4}. Red links with filled endpoints indicate $ZZ$ interaction gates \eqref{eq:Rzz} with the shown rotation angles. Green dots indicate single-qubit $Z$ rotations \eqref{eq:Rz} with the specified angles and $(\alpha_1,\alpha_2,\alpha_3,\alpha_4) =(-\frac{\pi}{4},-\frac{\pi}{2},-\pi,\frac{\pi}{4})$. Each of the six Givens gates $\hG_k$ comprises one $XY$ interaction and two single-qubit $Z$ rotations [Eq.~\eqref{eq:Givens-mb}], giving a total of six $XY$ and three $ZZ$ two-qubit gates.}
\end{figure}
For the special case of $n=2$ copies, $\hC_\A$ is simply a tensor product of $\hC=\SWAP$ operators for the sites of subsystem $\A$. Experimentally, it can be measured through the \emph{destructive SWAP test} by applying CNOT and Hadamard gates between the corresponding qubits of copy 1 and copy 2, followed by a measurement in the computational basis as
\begin{subequations}\label{eq:Cdiag-n2}
\begin{align}
	\SWAP&=\CNOT_{1,2}\,\Had_1\, \Omega\, \Had_1\, \CNOT_{1,2}\quad\text{with}\\
	\Omega&=\diag(1,1,1,-1)
\end{align}
\end{subequations}
containing the even and odd-parity eigenvalues $\omega_k$ of the SWAP operator \footnote{For $\CNOT_{i,j}$, qubit $i$ is the control bit, and qubit $j$ is the target bit.}.
The expectation value \eqref{eq:replica} is calculated directly from the parity of the resulting bit strings \cite{GarciaEscartin2013-87,Subasi2019-52}:
Let $K$ denote the number of (relevant) qubits in subsystem $\A$ \cite{Note1}. After applying $\Had_1\, \CNOT_{1,2}$ for each of these $K$ qubit pairs, measuring all $2K$ qubits in the $z$ basis yields $\omega_k=(-1)^{\nu_{k,1}\nu_{k,2}}$ in accordance with Eq.~\eqref{eq:Cdiag-n2}, and the expectation value over measurement shots gives the $n$-th moment
\begin{equation}\label{eq:expectC}
	\Tr(\dm_\A^n)=\EE\Big(\prod_{k=1}^K \omega_k\Big).
\end{equation}

Similar custom circuits which diagonalize the cyclic shift operators $\hC$ can be found for small $n$. Appendix~\ref{appx:small-n} provides a circuit for $n=3$ with six CNOT gates and a circuit for $n=4$ with 6 $XY$ and 3 $ZZ$ two-qubit gates; both circuits are shown in Fig.~\ref{fig:shiftTest-small-n}.

\subsection{Schemes for general \texorpdfstring{$n$}{n}}\label{sec:shiftTest-all-n}
To realize the shift-operator measurement for any number $n\geq 2$ of copies on a quantum computer, there are a number of options with different overheads:
\begin{enumerate}
 \item 
 One can proceed by adding one global ancilla qubit, applying a Hadamard gate to the ancilla and using it as the control qubit for the shift operator $\hC_\A$ acting across all $n$ copies, applying a second Hadamard, and measuring the ancilla. The probability $p_0$ of measuring $0$ on the ancilla yields the real part of the expectation value \eqref{eq:replica} with $\Tr(\dm_\A^n)=2p_0-1$. This approach of Johri et al.~\cite{Johri2017-96} has the drawback of requiring $\Theta(Kn)$ circuit depth.
 \item
 A very time-efficient scheme for measuring $\big(\Tr(\dm_\A^n)\big)^2$ with measurement-circuit depth $\O(1)$ is described by Suba\c{s}{\i} et al.~\cite{Subasi2019-52}. It is based on a two-copy test, requiring $2n$ copies of the causal-cone state and approximately doubling the qubit count. The scheme requires joint measurements on the relevant renormalized sites in both subsystems $\A$ and $\B$. These measurements can be performed as the sites leave the causal cones, allowing qubit reuse. In addition to the doubled number of required qubits, measuring the squared moment increases the sampling cost for a fixed entropy precision: For $C_n=\Tr(\dm_\A^n)\ll 1$ the required number of shots scales as $1/C_n^4$ in comparison to the scaling $1/C_n^2$ for the direct shift-operator measurements as discussed in Sec.~\ref{sec:samples}.
 \item
 As detailed in Secs.~\ref{sec:shiftTest-FFT} and \ref{sec:shiftTest-implement}, we suggest to employ a combination of a parity-controlled flux operator and a fermionic Fourier transform (FFT) to diagonalize the shift operator $\hC$ for each of the $K$ (relevant) qubits in subsystem $\A$ and its $n-1$ copies and to then measure all $Kn$ qubits in the computational basis. This approach requires only $\O(K n^2)$ two-qubit gates, $\O(n)$ circuit depth, and no qubit overhead.
\end{enumerate}
The method using the FFT and parity-controlled flux operator conserves the fermionic particle number (Hamming weight). Dropping that constraint, one can compile somewhat shorter measurement circuits, like the simple $\Had_1\, \CNOT_{1,2}$ circuit for $n=2$ in Sec.~\ref{sec:shiftTest-small-n}. Appendix~\ref{appx:small-n} provides such a custom circuit for $n=3$. It also describes a custom circuit for $n=4$, which employs the FFT but uses an alternative to the flux operator. Adaptive schemes using mid-circuit measurements would allow for further gate count reductions, and could be pursued in future work.

\subsection{Scheme using FFT and parity-controlled flux}\label{sec:shiftTest-FFT}
The cyclic shift operator \eqref{eq:Cprod} is unitary and hence diagonalizable. It is a tensor product with respect to the $K$ relevant subsystem-$\A$ qubits \cite{Note1} and, because
\begin{equation}\label{eq:C}
	\hC|\nu_1,\nu_2,\dotsc,\nu_n\ket = |\nu_n,\nu_1,\dotsc,\nu_{n-1}\ket
\end{equation}
such that $\hC^n=\id$, its eigenvalues are the $n$-th roots of unity
\begin{equation}\label{eq:Cspectrum}
	\omega_k\in\{1,\zeta,\zeta^2,\dotsc,\zeta^{n-1}\}\quad\text{with}\quad 
	\zeta=e^{\mri 2\pi/n},
\end{equation}
where $k=1,\dotsc,K$ labels the qubit.
In the following, let us consider one qubit of subsystem $\A$ and let $\hs^\alpha_j$ denote the Pauli operators for its copies $j=1,...,n$.

We can identify the computational basis states for this string of $n$ qubits with fermionic Fock states
\begin{equation}\label{eq:FockStates}
	|\nu_1,\nu_2,\dotsc,\nu_n\ket=(\ha_1^\dag)^{\nu_1}(\ha_2^\dag)^{\nu_2}\dots (\ha_n^\dag)^{\nu_n}|\text{vac}\ket
\end{equation}
of a corresponding $n$-site system of spinless fermions with occupation numbers $\nu_j\in\{0,1\}$.
According to the Jordan-Wigner transformation \cite{Jordan1928,Negele1988},
\begin{equation}\label{eq:JordanWigner}
	\ha^\dag_j  = \hs_j^-\prod_{i<j}\hs_i^z\quad\text{and}\quad
	\ha^\pdag_j = \hs_j^+\prod_{i<j}\hs_i^z
\end{equation}
are fermionic creation and annihilation operators with $\{\ha_i^\pdag,\ha_j^\dag\}=\delta_{i,j}$, $\hs^+=|0\ket\bra 1|$, and $\hs^z=|0\ket\bra 0|-|1\ket\bra 1|$.

The FFT $\hU_f$ is a unitary transformation for the fermionic single-particle basis, mapping from position eigenstates $|j\ket$ to quasimomentum eigenstates
\begin{equation}\label{eq:FFT-1particle}
	|\tilde{q}\ket=\sum_j [u_f]_{q,j}|j\ket
	\ \ \text{with}\ \
	[u_f]_{q,j}=\frac{1}{\sqrt{n}}\, e^{\mri \frac{2\pi}{n}\, qj}
\end{equation}
such that
\begin{equation}\label{eq:FFT}
	\ta_q^\dag:=\hU_f^\pdag\ha_q^\dag\hU_f^\dag=\sum_j [u_f]_{q,j}\ha_j^\dag,
\end{equation}
where $\ta_q^\dag$ and $\ta_q^\pdag$ are ladder operators for quasimomenta $q\,\frac{2\pi}{n}$ with $q=1,\dotsc,n$.
The FFT diagonalizes the \emph{fermionic} shift operator $\hC_f$ for $n$ spatial modes, where
\begin{equation}\label{eq:Cf}
	\hC_f\ha_j\hC_f^\dag=\ha_{j+1}
	\quad\text{and}\quad
	\hC_f = e^{-\mri \sum_q \frac{2\pi}{n}\,q\ta_q^\dag\ta_q^\pdag}.
\end{equation}

The only difference between the qubit shift operator $\hC$ and its fermionic counterpart $\hC_f$ is a sign factor due to
\begin{align}\nonumber
	&\hC_f|\nu_1,\nu_2,\dotsc,\nu_n\ket
	\stackrel{\eqref{eq:Cf}}{=}(\ha_2^\dag)^{\nu_1}(\ha_3^\dag)^{\nu_2}\dots (\ha_1^\dag)^{\nu_n}|\text{vac}\ket\\
	&\ =(-1)^{\nu_n\sum_{j=1}^{n-1}\nu_j}|\nu_n,\nu_1,\dotsc,\nu_{n-1}\ket,
	\label{eq:ShiftSign}
\end{align}
arising when we commute $(\ha_1^\dag)^{\nu_n}$ to the front, past all other creation operators. This sign factor can be realized by the parity-controlled flux operator
\begin{equation}\label{eq:F}
	\hF:=e^{\mri\frac{\pi}{n}\delta_{\hN,\text{even}}\sum_j j\hn_j}
	\ \ \text{with}\ \
	\hn_j=\ha_j^\dag\ha_j^\pdag=\frac{\id-\hs^z_j}{2}
\end{equation}
being the occupation number operators and $\hN:=\sum_{j=1}^n\hn_j$ the total particle number operator for the $n$ copies of the considered subsystem-$\A$ qubit, such that
\begin{subequations}
\begin{align}\label{eq:CFCfF}
	\hC &= e^{\mri\frac{\pi}{n}\hN\delta_{\hN,\text{even}}} \hF^\dag \hC_f \hF\\
	&= \hF^\dag \hU_f\cdot e^{\mri\frac{\pi}{n}\hN\delta_{\hN,\text{even}}} e^{-\mri \sum_j \frac{2\pi}{n}\,j\hn_j}\cdot \hU_f^\dag  \hF
	\label{eq:FUDUF}
\end{align}
\end{subequations}
The first equality is proven in Appendix~\ref{appx:F}. For the second equality, we have used expression \eqref{eq:Cf} for the fermionic shift operator, the relation \eqref{eq:FFT} between real-space and momentum-space ladder operators, and that $\hN$ commutes with the unitaries $\hF$ and $\hU_f$ as they conserve the particle number.

The diagonalized form~\eqref{eq:FUDUF} of the shift operator shows that we can measure the expectation value \eqref{eq:replica} of $\hC_\A$ as follows.
\begin{enumerate}
 \item
 Prepare $n$ copies of the state $|\Psi\ket$.
 \item
 For each of the $K$ relevant qubits of subsystem $\A$, apply the parity-controlled flux operator $\hF$ and the inverse FFT $\hU_f^\dag$.
 \item
 Measure all $Kn$ qubits in the computational basis, yielding occupation numbers $\nu_{k,j}\in\{0,1\}$ to obtain the associated $\hC$ eigenvalues
 \begin{equation}\label{eq:omega-k}
	\omega_k=e^{\mri\frac{\pi}{n}N_k\delta_{N_k,\text{even}}} e^{-\mri \sum_{j=1}^n \frac{2\pi}{n}\,j\nu_{k,j}}
 \end{equation}
 with $N_k=\sum_j\nu_{k,j}$, in accordance with Eq.~\eqref{eq:Cspectrum}.
 \item
 Averaging the associated $\hC_\A$ eigenvalues $\prod_{k=1}^K \omega_k$ over measurement shots according to Eq.~\eqref{eq:expectC}, we obtain $\Tr(\dm_\A^n)$.
\end{enumerate}
\begin{figure}[t]
	\includegraphics[width=\columnwidth]{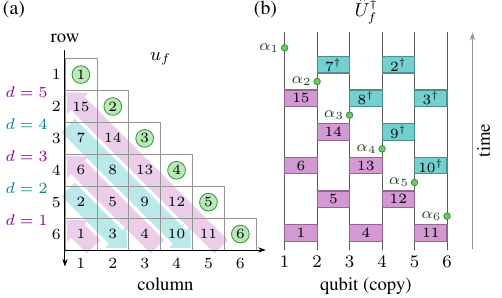}
	\caption{\label{fig:FFT}(a) We can use Givens rotations \cite{Golub1996} to sequentially nullify off-diagonal elements of the $n\times n$ single-particle basis-transformation matrix $u_f$ from Eq.~\eqref{eq:FFT-1particle} which defines the (many-body) fermionic Fourier transformation $\hU_f$ according to Eq.~\eqref{eq:FFT}. Proceeding in the specified order, the Clements scheme \cite{Clements2016-3} applies rotations to neighboring columns of the single-particle matrix when moving up-left along subdiagonals with odd index $d$ and applies rotations to neighboring rows of the single-particle matrix when moving down-right along subdiagonals with even index $d$.
	(b) On the many-body Hilbert space, these rotations correspond to a representation of $\hU_f^\dag$ as an $n$-layer brickwall circuit of $n(n-1)/2$ particle-number-preserving Givens rotations \eqref{eq:Givens-mb} as shown here for $n=6$ and in Fig.~\ref{fig:shiftTest-small-n}c for $n=4$. The remaining matrix elements on the main diagonal are implemented as single-qubit $Z$ rotations \eqref{eq:Rz}, indicated by the green dots.}
\end{figure}
\begin{figure}[t]
	\includegraphics[width=0.9\columnwidth]{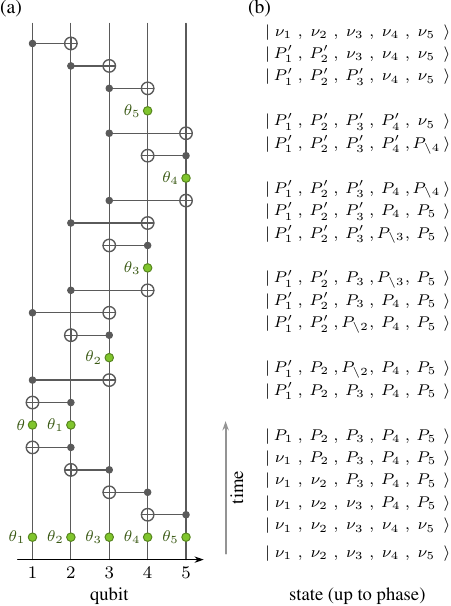}
	\caption{\label{fig:C}(a) The parity-controlled flux operator $\hF$ from Eq.~\eqref{eq:F} can be implemented as a circuit of $5n-8$ nearest and next-nearest neighbor CNOT gates, which encode certain partial parity quantum numbers in specific qubits on which we then act with single-qubit $Z$ rotations \eqref{eq:Rz} to implement $\hF$ ($2n+1$ rotations in total). See Appendix~\ref{appx:F-implement}. Panel (b) shows the associated sequence of intermediate states for $n=5$, when starting from a computational basis state $|\nu_1,\nu_2,\nu_3,\nu_4,\nu_5\ket$. $P_j=\nu_j\oplus\dotsb\oplus\nu_{n}$ and $P'_j=\nu_1\oplus\dotsb\oplus\nu_j$ are the right-partial and left-partial parities for qubit $j$, and $P_{\setminus j}=P'_j\oplus P_j$.}
\end{figure}

\subsection{Implementing FFT and parity-controlled flux}\label{sec:shiftTest-implement}
Finally, we need quantum circuits that implement the unitary parity-controlled flux operator $\hF$ and inverse FFT $\hU_f^\dag$.

Like any Bogoliubov transformation, the FFT $\hU_f$ can be decomposed into a circuit of $\O(n^2)$ Givens rotations (two-fermion gates) \cite{Wecker2015-92,Jiang2018-9,Kivlichan2018-120}: As detailed in Appendix~\ref{appx:Uf}, we can employ the Clements decomposition \cite{Clements2016-3} to implement $\hU_f$ as a brickwork circuit with $n$ layers, comprising a total of $n$ single-qubit $Z$ rotations and $n(n-1)/2$ particle-number-preserving Givens-rotation gates (beam splitters)
\begin{equation}\label{eq:Givens-mb}
	\hG_{i,j}(\theta, \phi) = e^{\mri \frac{2\phi+\pi}{4}\hs^z_i} \, \hR^{xy}_{i,j}(\theta) \, e^{-\mri \frac{2\phi+\pi}{4}\hs^z_i}
\end{equation}
with the $XY$ interaction gates $\hR^{xy}_{i,j}(\theta) = e^{-\mri \frac{\theta}{2} (\hs^x_i \hs^x_j + \hs^y_i \hs^y_j)}$ \cite{Makhlin1999-398,Imamoglu1999-83,Schuch2003-67,Steffen2006-313} acting on two neighboring qubits. The Givens rotations sequentially nullify off-diagonal elements of the single-particle basis transformation $u_f$ from Eq.~\eqref{eq:FFT-1particle}. Choosing them to act either on two neighboring rows or columns of the $n\times n$ matrix $u_f$ in the sequence shown in Fig.~\ref{fig:FFT}, Jordan-Wigner strings in their many-body representation cancel, and one obtains the simple two-qubit gates \eqref{eq:Givens-mb}. The remaining diagonal elements are implemented by $n$ single-qubit $Z$ rotations.

The parity-controlled flux operator $\hF$ from Eq.~\eqref{eq:F} is diagonal in the computational basis. As detailed in Appendix~\ref{appx:F-implement}, it can hence be implemented as a circuit of $5n-8$ nearest and next-nearest neighbor CNOT gates, which encode certain partial parity quantum numbers in specific qubits on which we then act with single-qubit $Z$ rotations to implement $\hF$ ($2n+1$ rotations in total). Figure~\ref{fig:C} shows this circuit and the associated sequence of intermediate states for $n=5$.

In this way, $\hF$ and $\hU_f^\dag$ are implemented by quantum circuits of depth $\Theta(n)$, containing a total of
\begin{subequations}\label{eq:shiftTest-gateCount}
\begin{alignat}{3}
	&n(n-1)/2\quad &&\text{two-qubit $XY$ gates \eqref{eq:Rxy}},\\
	&5n-8          &&\text{CNOT gates, and}\\
	&n^2+2n+1      &&\text{single-qubit $Z$ rotation gates,}
\end{alignat}
\end{subequations}
In total, we need
\begin{equation}\label{eq:cost-SC}
	n(n+9)/2-8\quad\text{CNOT and $XY$ gates}.
\end{equation}
This is, e.g., the relevant cost-measure for superconducting quantum computers with tunable qubits / couplers as built by Google and Rigetti, where a CNOT requires one controlled-phase gate, and $XY$ interaction (fSim) gates are native \cite{Arute2019-574,Foxen2020-125,Caldwell2018-10,Abrams2020-3}. (For simplicity, we ignore SWAP gates that may be necessary on superconducting quantum computers, depending on the qubit connectivity.)
The relevant cost-measure for trapped-ion quantum computers like IonQ Aria and Forte are the needed
\begin{equation}\label{eq:cost-ionTrap}
	n(n+4)-8\quad\text{CNOT or MS or $ZZ$ gates},
\end{equation}
where a CNOT is implemented with a single maximally entangling M\o{}lmer-S\o{}rensen (MS) or $ZZ$ gate, and an $XY$ interaction gate is realized using two MS or $ZZ$ gates \cite{Moelmer1999-82,Debnath2016-536,Maslov2017-19}. Alternatively, an $XY$ interaction gate can be implemented using two CNOT gates.

\section{Conclusion}\label{sec:conclusion}
We have developed replica-based schemes for the investigation of integer-order R\'{e}nyi entanglement entropies $S_n(\A)$ of MERA states on classical and quantum computers. Their central idea is the reduction to the causal cone of the subsystem boundary: Removing tensors that exclusively affect subsystem $\A$ or its complement $\B$ preserves the nonzero eigenvalues of the subsystem density-matrices $\dm_\A$ and $\dm_\B$ such that cyclic-shift measurements for the evaluation of moments $C_n=\Tr(\dm_\A^n)=\Tr(\hC_\A\dm_\A^{\otimes n})$ can be done directly on the renormalized sites that leave the causal cone into $\A$. The resulting tensor networks admit the classical contraction schemes and quantum implementations described in Secs.~\ref{sec:classical} and \ref{sec:quantum}. With midcircuit measurements and qubit reset, the 1D quantum scheme requires only $\Theta(n\log\chi)$ qubits. In 2D, the ordering of tensor applications allows for a tradeoff between circuit depth and quantum memory. By processing sequentially along the boundary, Scheme~B reduces the qubit count to $\Theta(n\log\ell\log\chi)$ compared to $\Theta(n\ell\log\chi)$ in Scheme~A and $\Theta(n\ell^2)$ in a full subsystem preparation as summarized in Table~\ref{tab:costs-cnct}.

The described cyclic-shift measurement circuits make this construction applicable for arbitrary integer R\'{e}nyi orders $n\geq2$. For each relevant subsystem-$\A$ qubit and its $n-1$ replicas, the fermionic Fourier transform and parity-controlled flux operator diagonalize the shift operator $\hC$ with $\O(n^2)$ two-qubit gates and $\O(n)$ circuit depth, without introducing any additional ancilla qubits. Custom circuits reduce the gate counts for the lowest orders $n=2,3$ and $4$, which are the most accessible. These circuits can also be useful for various applications outside the scope of this work. Their practical cost depends on the native entangling gates, qubit connectivity, and routing requirements as discussed in Sec.~\ref{sec:shiftTest}.

The total quantum computation time also depends on the statistical cost of resolving the entropy $S_n$ to a desired accuracy $\veps$. For a fixed failure probability and $0<(n-1)\veps\leq 1$, the analysis in Sec.~\ref{sec:samples} gives the sufficient shot budget $\Nshots=\O\big(e^{2(n-1)S_n}/[(n-1)^2\veps^2]\big)$. At fixed $n$ and $\veps$, this bound is independent of subsystem size for 1D area-law states and grows polynomially with $\ell$ for 1D critical states described by conformal field theories. For higher-dimensional area-law and log-area-law states, the bound is exponential in $\ell^{D-1}$ and $\ell^{D-1}\log\ell$, respectively, which raises the question whether the quantum implementations for $D\geq 2$ are usually limited to small subsystems.

Section~\ref{sec:samples-tomog} discusses alternative (holographic) approaches using tomography or classical shadows and independent uniform local Pauli measurements. While they do not require replicas, the sufficient shot budgets obtained in Appendices~\ref{appx:tomog} and \ref{appx:shadow} feature a multiplicative overhead which scales exponentially in the number $K$ of relevant subsystem-$\A$ qubits. Moreover, for a family of ternary MERA, we found that estimating the entropy requires $\Omega\big((3/2)^K/\veps^2\big)$ single-copy Pauli measurements, whereas replica measurements need only $\O(1/\veps^2)$ shots at fixed $n$ and failure probability. This exponential separation in $K$ holds for arbitrary classical processing of the measurement records.

In experiments, the achievable accuracy additionally depends on state-preparation errors, imperfect cyclic-shift gates, and readout errors. Equation~\eqref{eq:Cn-precision} provides a criterion for error budget allocation in terms of the sufficient moment-accuracy condition $|C_n^\est-C_n|\leq\delta_n=\frac{1}{2}(n-1)\veps C_n$. The companion study Ref.~\cite{Jeanette2026_12} details error mitigation strategies and demonstrates the successful measurement of R\'{e}nyi-2 and R\'{e}nyi-3 entanglement entropies on a trapped-ion quantum computer, resolving logarithmic growth and area-law saturation in critical and gapped spin chains.

Further developments could optimize the shift-operator measurements for specific hardware, including adaptive circuits that use intermediate measurements. For systems in $D\geq 2$ spatial dimensions, an important question is whether additional structure of the states and observables can be used to further reduce the sampling cost for small moments and extend the range of entanglement scales accessible in quantum simulations.

\begin{acknowledgments}
I gratefully acknowledge discussions with Liam Jeanette, Qiang Miao, and Norbert M.\ Linke.
\end{acknowledgments}

\appendix

\newpage
\section{Sample complexity for Pauli full-state tomography}\label{appx:tomog}
This appendix provides a proof for the bound \eqref{eq:Nshots-tomog} on the number of samples needed to determine the R\'{e}nyi-$n$ entanglement entropy \eqref{eq:Renyi} to precision $\veps$ [Eq.~\eqref{eq:Sn-precision}] with fixed success probability through full state tomography on the $K$ relevant qubits in subsystem $\A$ \cite{Note1}.

As described in Sec.~\ref{sec:samples-tomog}, $\Nshots$ Pauli measurements of the $K$ qubits yield the sample average
\begin{equation}
	\bar{\rho} \equiv \Bra\dm_s\Kets \equiv \frac{1}{\Nshots}\sum_s\dm_s
\end{equation}
with snapshot operators $\dm_s$ as defined in Eq.~\eqref{eq:PauliSnapshot}, and
\begin{equation}
	\dm_\est=\argmin_{\dm\succeq 0,\, \Tr\dm=1}\|\dm-\bar{\rho}\|_2
\end{equation}
is the Frobenius-norm projection of the sample average $\bar{\rho}$ onto the closed convex set of density operators.

\subsection{Reconstruction accuracy}
Trace norm distances can be bounded by Frobenius-norm distances as $\|\Delta\dm\|_1\leq\sqrt{2^K}\|\Delta\dm\|_2$ such that
\begin{equation}\label{eq:tomog-1}
	\EE\|\dm_\est-\dm_\A\|_1^2
	\leq 2^K \EE\|\dm_\est-\dm_\A\|_2^2
	\leq 2^K \EE\|\bar{\rho}-\dm_\A\|_2^2.
\end{equation}
In the second step, we have used that the projection $\bar{\rho}\to\dm_\est$ cannot increase the Frobenius-norm distance to $\dm_\A$.
Defining the zero-mean fluctuations $\Delta\dm_s:=\dm_s-\dm_\A$, we have
\begin{align}\label{eq:tomog-2}
	\EE\|\bar{\rho}-\dm_\A\|_2^2
	&=\frac{1}{\Nshots^2}\sum_{s,r}\EE\Tr(\Delta\dm_s\Delta\dm_r)\\
	&=\frac{1}{\Nshots^2}\sum_{s}\EE\Tr(\Delta\dm_s^2)
	=\frac{1}{\Nshots}\EE\|\dm_s-\dm_\A\|_2^2.\nonumber
\end{align}
With $\Tr(\dm_s^2)=5^K$, Eqs.~\eqref{eq:tomog-1} and \eqref{eq:tomog-2} give the bound
\begin{equation}\label{eq:tomog-3}
	\EE\|\dm_\est-\dm_\A\|_1^2 \leq \frac{10^K}{\Nshots}.
\end{equation}
Then, application of Markov's inequality $\Pr(X\geq \eta)\leq \EE(X)/\eta$ to Eq.~\eqref{eq:tomog-3} establishes that state tomography with Pauli measurements can achieve trace-norm accuracy $\|\dm_\est-\dm_\A\|_1\leq\eta$ with any desired fixed probability by scaling the number of shots $\Nshots^\est(\eta)$ as \cite{Yu2020_09}
\begin{equation}\label{eq:Nshots-est}
	\Nshots^\est(\eta) = \O\left(\frac{10^K}{\eta^2}\right).
\end{equation}

\subsection{R\'{e}nyi accuracy}
We can now use the reconstruction accuracy $\eta$ to bound the deviation $\Delta C_n:=|\Tr(\dm_\est^n-\dm_\A^n)|$ of the $n$-th moment estimator in Eq.~\eqref{eq:Cn-tomog}. Writing the difference as a telescoping sum $\dm_\est^n-\dm_\A^n=\sum_{m=0}^{n-1}\dm_\est^{n-1-m}(\dm_\est - \dm_\A)\dm_\A^m$ and employing the submultiplicative property
\begin{equation}\label{eq:traceNorm-sub}
	\|\hat{A}\hat{B}\|_1\leq\|\hat{A}\|\|\hat{B}\|_1
\end{equation}
of the trace norm with respect to the operator norm as well as $\|\dm_\est^m\|\leq 1$ and $\|\dm_\A^m\|\leq 1$, we have
\begin{align}
	\Delta C_n
	&\leq \sum_{m=0}^{n-1}\left|\Tr\left(\dm_\est^{n-1-m}(\dm_\est - \dm_\A)\dm_\A^m\right)\right| \nonumber\\
	&\leq \sum_{m=0}^{n-1}\left\|\dm_\est^{n-1-m}(\dm_\est - \dm_\A)\dm_\A^m\right\|_1 \nonumber\\
	&\leq n\|\dm_\est - \dm_\A\|_1\leq n\eta
\end{align}

Thus, choosing the reconstruction accuracy bound
\begin{equation}
	\eta = \frac{\delta_n}{n} \stackrel{\eqref{eq:Cn-precision}}{=} \frac{(n-1)\veps C_n}{2n},
\end{equation}
we achieve the sufficient moment accuracy \eqref{eq:Cn-precision} and the desired entropy precision $\veps$ in Eq.~\eqref{eq:Sn-precision} with
\begin{equation*}
	\Nshots^\text{tomo}
	\!\stackrel{\eqref{eq:Nshots-est}}{=}\!
	 \O\!\left( \frac{10^K n^2}{(n-1)^2\veps^2C_n^2} \right)
	\!\stackrel{\eqref{eq:Renyi}}{=}
	 \O\!\left( \frac{10^K n^2 e^{2(n-1)S_n}}{(n-1)^2\veps^2} \right).
\end{equation*}

\section{Sample complexity for Pauli classical shadows}\label{appx:shadow}
This appendix proves the bound \eqref{eq:Nshots-shadow} on the number of samples needed to determine the R\'{e}nyi-$n$ entanglement entropy \eqref{eq:Renyi} to precision $\veps$ [Eq.~\eqref{eq:Sn-precision}] with fixed success probability using classical shadows based on Pauli measurements for the $K$ relevant qubits in subsystem $\A$ \cite{Note1}.

As described in Sec.~\ref{sec:samples-tomog}, the $\Nshots$ measurement results give the classical snapshots
\begin{equation}\label{eq:PauliSnapshot2}
	\dm_s=\bigotimes_{k=1}^K \dml_{s,k},\ \ \text{where}\ \
	\dml_{s,k}=\frac{1}{2}\left(\id+3m_{s,k}\hs^{\alpha_{s,k}}\right)
\end{equation}
with measurement axes $\alpha_{s,k}$ and measurement outcomes $m_{s,k}=\pm 1$ for sample $s$ and qubit $k$.

As $\EE(\dm_s)=\dm_\A$, the real-valued moment estimator $C^\est_n$ defined in Eq.~\eqref{eq:Cn-shadow} obeys
\begin{equation}
	\EE(C^\est_n) = \Re\Tr\big(\EE(\dm_s)^n\big) = \Tr(\dm_\A^n) \equiv C_n.
\end{equation}
The exclusion of repeated sample indices in the sum of Eq.~\eqref{eq:Cn-shadow} is essential for this unbiasedness.

\subsection{Moment accuracy}
Defining for every ordered tuple $\vs=(s_1,\dotsc,s_n)$ of distinct sample indices $s_j\in\{1,\dotsc,\Nshots\}$ in the estimator \eqref{eq:Cn-shadow} the summand
\begin{equation}\label{eq:shadow-cs}
	c_\vs:=\Re\Tr(\dm_{s_1}\dots\dm_{s_n})
	=\Re\prod_{k=1}^K \!\Tr(\dml_{s_1,k}\dots\dml_{s_n,k})
\end{equation}
and the zero-mean deviation $d_\vs:=c_\vs-C_n$, the mean squared deviation of the estimator $C^\est_n$ from the exact moment $C_n$ is
\begin{equation}\label{eq:shadow-1}
	\EE|C^\est_n-C_n|^2
	= \left(\frac{(\Nshots-n)!}{\Nshots!}\right)^2 \sum_{\vs,\vs'}\EE(d_\vs d_{\vs'}).
\end{equation}
The contribution of a pair of tuples $\vs$ and $\vs'$ in Eq.~\eqref{eq:shadow-1} depends on the number $r$ of sample indices they share. For $r=0$, independence and $\EE(d_\vs)=0$ imply $\EE(d_\vs d_{\vs'})=0$.

For $1\leq r<n$, let us condition the two summands $c_\vs$ and $c_{\vs'}$ on the $r$ shared snapshots and label them $\dm_{s_1},\dotsc,\dm_{s_r}$, while retaining their original locations in the products. The two summands are then conditionally independent, and their conditional expectations $c_{\vs,r}:=\EE(c_\vs|\dm_{s_1},\dotsc,\dm_{s_r})$ and $c_{{\vs'},r}$ are obtained by replacing each unshared snapshot by $\dm_\A$. Applying the submultiplicative property \eqref{eq:traceNorm-sub} of the trace norm gives
\begin{equation}\label{eq:shadows-conditional-bound}
	|c_{\vs,r}|,\,|c_{\vs',r}| \leq\|\dm_\A\|_1 \|\dm_\A\|^{n-r-1} \prod_{j=1}^{r}\|\dm_{s_j}\| \leq 2^{rK},
\end{equation}
where we have used $\|\dm_\A\|_1=1$, $\|\dm_\A\|\leq 1$, and $\|\dm_s\|=2^K$.
Since $\EE(c_{\vs,r})=\EE(c_{\vs',r})=C_n$, the Cauchy-Schwarz inequality gives
\begin{align}\label{eq:shadow-pair}
	|\EE&(d_\vs d_{\vs'})|= |\EE\big((c_{\vs,r}-C_n)(c_{\vs',r}-C_n)\big)|\nonumber\\
	&\leq \sqrt{\big(\EE(c_{\vs,r}^2)-C_n^2\big)\big(\EE(c_{\vs',r}^2)-C_n^2\big)}
	\leq 4^{rK}.
\end{align}

To see that the same bound also follows for $r=n$, consider a general single-qubit operator $\hR=R_0\id+\sum_\alpha R_\alpha\hs^\alpha$ and write $\dml(\alpha,m):=(\id+3m\hs^\alpha)/2$. Then
\begin{align}\label{eq:shadow-local}
	&\frac{1}{3}\sum_{\alpha=x,y,z}\max_{m=\pm 1} \big|\Tr\big(\hR\dml(\alpha,m)\big)\big|^2\nonumber\\
	&\quad\leq |R_0|^2+3\sum_\alpha|R_\alpha|^2 +2|R_0|\sum_\alpha|R_\alpha|\nonumber\\
	&\quad\leq 4\Big(|R_0|^2+\sum_\alpha|R_\alpha|^2\Big) = 2\|\hR\|_2^2,
\end{align}
where the second inequality uses $2|R_0||R_\alpha|\leq |R_0|^2+|R_\alpha|^2$.
For the first $n-1$ snapshots in $c_\vs$, define $\hR_k:=\dml_{s_1,k}\dots\dml_{s_{n-1},k}$. Submultiplicativity of the Frobenius norm with respect to the operator norm and $\|\dml_{s,k}\|_2^2=5$ give
\begin{equation}\label{eq:shadow-product}
	\|\hR_k\|_2^2 \leq \|\dml_{s_1,k}\|_2^2 \prod_{j=2}^{n-1}\|\dml_{s_j,k}\|^2 = 5\cdot 4^{n-2}.
\end{equation}
The trace of the snapshot product factorizes over qubits as in Eq.~\eqref{eq:shadow-cs}. Maximizing independently over the outcomes of the last sample before averaging over its randomly chosen measurement axes, Eqs.~\eqref{eq:shadow-local} and \eqref{eq:shadow-product} imply
\begin{equation*}
	\EE\big|\!\Tr(\dm_{s_1}\dots\dm_{s_n})\big|^2
	\leq \EE\prod_{k=1}^K 2\|\hR_k\|_2^2 \leq (10\cdot 4^{n-2})^K \leq 4^{nK}.
\end{equation*}
Taking real parts cannot increase the second moment, so $\EE(d_\vs^2)=\EE(c_\vs^2)-C_n^2\leq4^{nK}$, and the Cauchy-Schwarz inequality establishes Eq.~\eqref{eq:shadow-pair} also for $r=n$, including tuples that contain the same sample indices in different orders.

It remains to count the pairs of tuples $(\vs,\vs')$ with $r$ shared sample indices. For $\Nshots\geq 2n$ samples, their fraction is
\begin{equation}\label{eq:shadow-overlap}
	p_r = \frac{\binom{n}{r}\binom{\Nshots-n}{n-r}}{\binom{\Nshots}{n}}
	\leq \binom{n}{r}\left(\frac{2n}{\Nshots}\right)^r
\end{equation}
as there are a total of $\binom{\Nshots}{n}$ possible index sets for the second tuple, we have $\binom{n}{r}$ choices for the $r$ shared indices of the first tuple's $n$ indices, and $\binom{\Nshots-n}{n-r}$ ways to choose the remaining $n-r$ indices of the second tuple from the $\Nshots-n$ indices outside the first tuple. For the upper bound, sum over the $\binom nr$ choices of $r$ indices in the first tuple and bound the probability that the second tuple contains a specified choice by $\prod_{j=0}^{r-1}(n-j)/(\Nshots-j)\leq(2n/\Nshots)^r$.

In conjunction, Eqs.~\eqref{eq:shadow-1}, \eqref{eq:shadow-pair}, and \eqref{eq:shadow-overlap} give
\begin{align}\label{eq:shadow-2}
	\EE|C^\est_n-C_n|^2
	&\leq \sum_{r=1}^n p_r 4^{rK}
	\leq \left(1+\frac{2n 4^K}{\Nshots}\right)^n-1\nonumber\\
	&\leq \exp\left(\frac{2n^2 4^K}{\Nshots}\right)-1,
\end{align}
where the second step uses the binomial theorem. For $\Nshots\geq2n^2 4^K$, we can use $e^x-1\leq2x$ for $0\leq x\leq1$ to arrive at the bound
\begin{equation}\label{eq:shadow-3}
	\EE|C^\est_n-C_n|^2\leq\frac{4n^2 4^K}{\Nshots}
\end{equation}
and Markov's inequality establishes
\begin{equation}\label{eq:shadow-Markov}
	\Pr(|C^\est_n-C_n|\geq\delta_n)
	\leq\frac{\EE|C^\est_n-C_n|^2}{\delta_n^2}
	\leq\frac{4n^2 4^K}{\Nshots\delta_n^2}.
\end{equation}
Thus, choosing $\Nshots \geq 4n^2 4^K/(p_\err\delta_n^2)$ with $0 < \delta_n \leq 1$, Pauli classical shadows achieve moment accuracy $|C^\est_n-C_n|\leq\delta_n$ with probability at least $1-p_\err$, and we also have $\Nshots\geq2n^2 4^K$. At fixed $p_\err$, the sufficient sample count hence scales as
\begin{equation}\label{eq:Nshots-shadow-moment}
	\Nshots^\text{mom}(\delta_n) = \O\left(\frac{n^2 4^K}{\delta_n^2}\right).
\end{equation}

\subsection{R\'{e}nyi accuracy}
\begin{figure*}[t]
	\centering
	\includegraphics[width=0.92\textwidth]{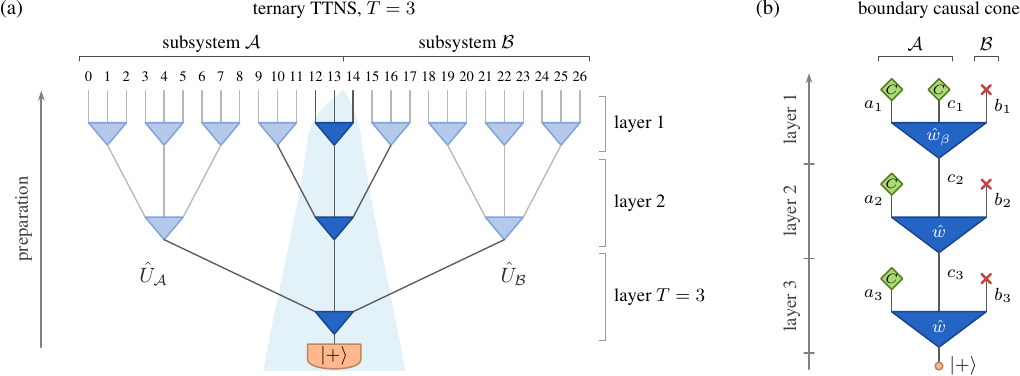}
	\caption{\label{fig:TTNSternary}The family of ternary TTNS considered in Appendix~\ref{appx:PauliLower-MERA} for bond dimension $\chi=2$, $T=3$ layers, and $3^T=27$ sites with open boundary conditions. (a) With subsystem $\A$ containing the first $\ell=(3^T+1)/2=14$ sites, the causal cone of the subsystem boundary is formed by the central trunk of tensors (unshaded). All shaded isometries may be chosen arbitrarily \cite{Note3} and are removed by the subsystem-local inverse basis transformations $\hU_\A^\dag$ and $\hU_\B^\dag$. (b) The retained causal-cone network with the isometry $\hw$ [Eq.~\eqref{eq:PauliLower-copy}] in layers $2,3,\dotsc$ and $\hw_\beta$ [Eq.~\eqref{eq:PauliLower-final}] in layer $1$. Tracing out subsystem $\B$ (red crosses) produces the reduced state $\dm_{\A,\beta}=\frac{1}{2}\big(|0^K\ket\bra0^K|+|1^K\ket\bra1^K|+\beta|0^K\ket\bra1^K|+\beta|1^K\ket\bra0^K|\big)$ [Eq.~\eqref{eq:PauliLower-state}] on the $K=T+1=4$ relevant subsystem-$\A$ qubits $a_3,a_2,a_1,c_1$. Measuring the cyclic shift operator $\hC^{\otimes K}$ on these sites (green diamonds) yields the $n$-th moment $C_n=\Tr(\dm_{\A,\beta}^n)$ [Eq.~\eqref{eq:PauliLower-moment}]. As in Fig.~\ref{fig:measure}, each tensor represents $n$ copies according to the replica trick.}
\end{figure*}
According to the discussion in Sec.~\ref{sec:samples}, the estimator $S^\est_n$ from Eq.~\eqref{eq:Cn-clipped} achieves R\'{e}nyi entropy accuracy $\veps$ for moment tolerance $\delta_n=(n-1)\veps C_n/2$.

Thus, classical shadows achieve entropy accuracy $\Delta S_n\leq\veps$ when scaling the number of samples as
\begin{equation*}
	\Nshots^\text{shadow}
	\!\stackrel{\eqref{eq:Nshots-shadow-moment}}{=}
	 \O\!\left(\frac{4^K n^2}{(n-1)^2\veps^2C_n^2}\right)
	\!\stackrel{\eqref{eq:Renyi}}{=}
	 \O\!\left(\frac{4^K n^2 e^{2(n-1)S_n}} {(n-1)^2\veps^2}\right)
\end{equation*}
and $\Nshots^\text{shadow} \geq 2n^2 4^K$.

\section{Lower bound for uniform Pauli measurements}\label{appx:PauliLower}
This appendix proves the lower bound \eqref{eq:Nshots-PauliLower} on the sample complexity for a family of MERA when the R\'{e}nyi entropy estimator is given independent Pauli measurement records \eqref{eq:PauliSnapshot} for the holographic subsystem state \eqref{eq:PauliLower-state}. This covers both state tomography and classical shadows as discussed in Appendices~\ref{appx:tomog} and \ref{appx:shadow}. More generally, the estimator may know the considered state family, but not its parameter $\beta$. We require entropy accuracy $\veps$ with probability at least $1-p_\err$ uniformly over the interval $\beta\in[1/3,2/3]$. The R\'{e}nyi order $n\geq2$ and $p_\err$ are fixed, and constants in the complexity bounds may depend on them.

\subsection{The family of ternary MERA}\label{appx:PauliLower-MERA}
Consider a ternary MERA with bond dimension $\chi=2$ and $T\geq 1$ layers for a spin-$1/2$ chain of $3^T$ sites with open boundary conditions. All disentanglers shall be trivial (be the identity) such that we deal with a tree tensor network state (TTNS) \cite{Shi2006-74,Tagliacozzo2009-80,Barthel2026-9}. Let subsystem $\A$ consist of the first $\ell=(3^T+1)/2$ sites. The causal cone of the subsystem boundary then consists of the central trunk of ternary isometries. The left and central outgoing legs of the layer-1 isometry map to the two right-most sites of subsystem $\A$, and the right outgoing leg maps to the left-most site of subsystem $\B$.

As illustrated in Fig.~\ref{fig:TTNSternary}, the root is initialized in the state $|+\ket=\frac{1}{\sqrt{2}}(|0\ket+|1\ket)$, and we use
\begin{equation}\label{eq:PauliLower-copy}
	\hw|\nu\ket=|\nu\ket_a\otimes|\nu\ket_c\otimes|0\ket_b
	\quad\text{with}\quad \nu\in\{0,1\}
\end{equation}
for the isometries of layers $2,\dotsc,T$ in the boundary causal cone. The left child $a$ leaves the causal cone into subsystem $\A$, and the right child $b$ leaves into subsystem $\B$. The layer-1 isometry is
\begin{equation}\label{eq:PauliLower-final}
	\hw_\beta|\nu\ket=|\nu\ket_a\otimes|\nu\ket_c\otimes|f_{\beta,\nu}\ket_b
\end{equation}
with
\begin{equation}\label{eq:PauliLower-environment}
	|f_{\beta,\nu}\ket: =e^{\mri\frac{\theta}{2}(-1)^\nu\hs^z}|+\ket\quad \text{and} \quad
	\theta=\arccos\beta.
\end{equation}
On the physical layer, the outputs $a$ and $c$ of the isometry \eqref{eq:PauliLower-final} belong to $\A$, while $b$ belongs to $\B$. All isometries outside the boundary causal cone can be chosen arbitrarily \footnote{In fact, we could allow for MERA structure outside the boundary causal cone, i.e., also include unitary (dis)entanglers in those parts of the network. All results of Appendix~\ref{appx:PauliLower} also apply to this larger family of states and result in the lower complexity bound \eqref{eq:Nshots-PauliLower}.}.

Applying the inverse basis transformations $\hU_\A^\dag$ and $\hU_\B^\dag$ in the two subsystems removes all isometries outside the causal cone of the subsystem boundary \cite{Note1} as illustrated in Figs.~\ref{fig:MERA} and \ref{fig:TTNSternary}b. Omitting the decoupled ancillas, this leaves us with the boundary causal-cone state
\begin{equation}\label{eq:PauliLower-purification}
	|\Psi_\beta\ket =\frac{1}{\sqrt{2}}\sum_{\nu=0,1}|\nu\ket_\A^{\otimes K}\otimes|f_{\beta,\nu}\ket_b\otimes|0\ket_\B^{\otimes(T-1)},
\end{equation}
where $K=T+1$. Since $\bra f_{\beta,1}|f_{\beta,0}\ket=\cos\theta=\beta$, tracing out site $b$ and the remaining $T-1$ renormalized sites of subsystem $\B$ gives the density operator $\dm_{\A,\beta}$ for the $K$ relevant (renormalized) sites of subsystem $\A$ \cite{Note1} as stated in Eq.~\eqref{eq:PauliLower-state}. Its nonzero eigenvalues coincide with those of the full physical subsystem density operator, and
\begin{equation}\label{eq:PauliLower-moment}
	C_n:=\Tr(\dm_{\A,\beta}^n)=\lambda_+^n+\lambda_-^n\ \ \text{with}\ \
	\lambda_\pm=\frac{1\pm \beta}{2}
\end{equation}
such that $C_n\geq2^{1-n}$ and $0\leq S_n\leq\log 2$ for all $\beta$ and numbers of layers $T$.

\subsection{Information in the Pauli measurement records}
Let $\valpha=(\alpha_1,\dotsc,\alpha_K)$ and $\vm=(m_1,\dotsc,m_K)$ denote the axes and outcomes for a Pauli measurement as described in Sec.~\ref{sec:samples-tomog}. With the subsystem density operator \eqref{eq:PauliLower-state} and fixed axes $\valpha$, outcome $\vm$ occurs with probability
\begin{equation}\label{eq:PauliLower-probability}
	p_\beta(\vm|\valpha) = \Tr\Big(\dm_{\A,\beta}\bigotimes_{k=1}^K\frac{1}{2}\,(\id+m_k\hs^{\alpha_k})\Big).
\end{equation}
If any axis $\alpha_k=z$, the corresponding projector has vanishing matrix elements between $|0\ket_k$ and $|1\ket_k$ such that the $\beta$-dependent terms
drop out of Eq.~\eqref{eq:PauliLower-probability}. In contrast, if every axis $\alpha_k$ is either $x$ or $y$, then
\begin{align}\nonumber
	p_\beta(\vm|\valpha)
	&= \frac{1}{2^K}\Big( 1+\beta\cos(N_y\pi/2)\prod_k m_k \Big)\\
	&= \frac{1}{2^K}\Big( 1+\beta\delta_{N_y,\text{even}}\Pi \Big)\
	\label{eq:PauliLower-xy},
\end{align}
where $N_y$ is the number of $y$ axes and $\Pi:=(-1)^{N_y/2}\prod_k m_k$. Thus, $p_\beta(\vm|\valpha)$ contains information about $\beta$ only if all axes are $x$ or $y$ and $N_y$ is even. The fraction of such informative settings among the $3^K$ equally likely Pauli axes choices is
\begin{equation}\label{eq:PauliLower-informative}
	q = \frac{2^{K-1}}{3^K}=\frac{1}{2}\left(\frac{2}{3}\right)^K.
\end{equation}

Let $P_\beta(\valpha,\vm):=\frac{1}{3^{K}}\,p_\beta(\vm|\valpha)$ denote the joint distribution of axis orientations $\valpha$ and outcomes $\vm$, and consider two MERA parameters $\beta$ and $\beta'$. As $P_\beta(\valpha,\vm)=P_{\beta'}(\valpha,\vm)$ for uninformative settings and $\Pr(\Pi=\pm 1)=\frac{1}{2}(1\pm \beta)$ for informative settings, the relative entropy (a.k.a.\ Kullback–Leibler divergence) of the two distributions is
\begin{align}\nonumber
	D(P_\beta\|P_{\beta'})
	&\equiv \sum_{\valpha,\vm} P_\beta(\valpha,\vm)\log\frac{P_\beta(\valpha,\vm)}{P_{\beta'}(\valpha,\vm)}\\
	&=q\sum_{\Pi=\pm 1}\frac{1+\Pi\beta}{2}\log\frac{1+\Pi\beta}{1+\Pi\beta'}\nonumber\\
	&\leq q\frac{(\beta-\beta')^2}{1-\beta^{\prime2}} \leq\frac{9q}{5}\,(\beta-\beta')^2,
	\label{eq:PauliLower-KL}
\end{align}
where the first inequality follows from $\log x\leq x-1$ and the second assumes $\beta,\beta'\in[1/3,2/3]$.

\subsection{Lower bound on the number of shots}
According to Eq.~\eqref{eq:PauliLower-moment}, the derivative $\frac{\ud}{\ud\beta} S_n$ of the R\'{e}nyi-$n$ subsystem entropy \eqref{eq:Renyi} is continuous and strictly negative for $\beta\in[1/3,2/3]$. Let $c_n:=\min_\beta|\frac{\ud}{\ud\beta} S_n|$ on this interval and choose
\begin{equation}\label{eq:PauliLower-two-point}
	\beta=\frac{1}{2}\quad\text{and}\quad\beta'=\frac{1}{2}+\frac{4\veps}{c_n}
\end{equation}
for an $\veps>0$ that is small enough such that $\beta,\beta'\in[1/3,2/3]$. The two associated entropies then differ by at least $4\veps$.

An entropy estimator with error at most $\veps$ and success probability $\geq 1-p_\err$ for both parameters therefore defines a test that distinguishes $P^{\otimes\Nshots}_\beta$ and $P^{\otimes\Nshots}_{\beta'}$ with error probability at most $p_\err$.

For two distributions $P,Q$, the smallest average error probability in binary hypothesis testing with equal prior probabilities is
\begin{equation}
	\frac{1}{2}\sum_z\min\{P(z),Q(z)\}=\frac{1}{2}-\frac{1}{4}\|P-Q\|_1,
\end{equation}
where $\|P-Q\|_1=\sum_z|P(z)-Q(z)|$ is the $L_1$ norm distance \cite{Tsybakov2009}. The entropy-based test just described thus requires $\|P^{\otimes\Nshots}_\beta-P_{\beta'}^{\otimes\Nshots}\|_1\geq 2-4p_\err$. Pinsker's inequality, additivity of the relative entropy for product distributions, and Eq.~\eqref{eq:PauliLower-KL} give
\begin{align}\label{eq:PauliLower-testing}
	4(1&-2p_\err)^2 \leq \big\|P^{\otimes\Nshots}_\beta-P^{\otimes\Nshots}_{\beta'}\big\|_1^2
	\leq 2 D\big(P_\beta^{\otimes\Nshots}\big\|P_{\beta'}^{\otimes\Nshots}\big) \nonumber\\
	&=2\Nshots D(P_\beta\|P_{\beta'})
	\leq\frac{18 q\Nshots}{5}\,(\beta-\beta')^2.
\end{align}

With Eqs.~\eqref{eq:PauliLower-informative} and \eqref{eq:PauliLower-two-point} we obtain the necessary sample count
\begin{equation}\label{eq:PauliLower-explicit}
	\Nshots \geq \frac{5c_n^2(1-2p_\err)^2}{36\,\veps^2}\left(\frac{3}{2}\right)^K,
\end{equation}
which proves Eq.~\eqref{eq:Nshots-PauliLower}. This argument holds for arbitrary classical postprocessing; no assumption about the specific form of the estimator entered.

\subsection{Comparison with replica-based measurements}
For the considered family of MERA, we found $S_n\leq\log 2$ in Sec.~\ref{appx:PauliLower-MERA} for all $\beta$ and $K$. The fixed-confidence bound in Sec.~\ref{sec:samples} therefore implies that replica measurements achieve entropy accuracy $\veps$ with
\begin{equation}\label{eq:PauliLower-replica-shots}
	\Nshots^\text{replica} = \O\left(\frac{4^{n-1}}{(n-1)^2\veps^2}\right) = \O\left(\frac{1}{\veps^2}\right)
\end{equation}
for fixed $n$ and $p_\err$, independently of $K$.

With circuit depth $\Theta(T)=\Theta(K)$, Eqs.~\eqref{eq:PauliLower-explicit} and \eqref{eq:PauliLower-replica-shots} imply execution times
\begin{equation*}
	\tau^\text{replica}=\O\left(\frac{\log\ell}{\veps^2}\right)\ \ \text{and}\ \
	\tau^\text{Pauli}=\Omega\left(\frac{\ell^{\log_3(3/2)}\log\ell}{\veps^2}\right)
\end{equation*}
as $K=T+1$ and $\ell=(3^T+1)/2$.

\subsection{Extensions of the state family}
The restriction to uniform Pauli measurements in subsystem $\A$ matters for the family of ternary MERA considered above: Measuring all $K$ subsystem-$\A$ qubits along the $x$ axis (which is exponentially unlikely for a uniformly at random axis choice) would give direct access to $\beta$. Also, according to Eq.~\eqref{eq:PauliLower-purification}, the reduced state of the left-most subsystem-$\B$ qubit is $(\id+\beta\hs^x)/2$, i.e., gives local access to $\beta$.

Both shortcuts can be removed by modifying the family of states: Use an even number of layers $T=2T'\geq 4$. In preparation order, construct the central tensor trunk out of $T'-1$ copies of $\hw_\A:=\hw$ from Eq.~\eqref{eq:PauliLower-copy}, followed by $\hw_\beta^\text{mid}$, then $T'-1$ copies of $\hw_\B$, and finally $\hw^\text{end}$, where
\begin{align*}
	\hw_\beta^\text{mid}|\nu\ket&=|\nu\ket_a\otimes|f_{\beta,\nu}\ket_c\otimes|0\ket_b,\\
	\hw_\B|\nu\ket&=|0,\nu,\nu\ket,\quad\text{and} \quad
	\hw^\text{end}|\nu\ket=|0,0,\nu\ket.
\end{align*}
There are now $T'$ nontrivial outgoing qubits leaving the causal cone into $\A$ and $T'$ leaving into $\B$. Additionally, we apply a single-qubit unitary $\hat{V}$ to each corresponding index of the trunk's isometries with $\hat{V}\hs^z\hat{V}^\dag=(\hs^x+\hs^y+\hs^z)/{\sqrt{3}}$. All isometries outside the boundary causal cone may again be arbitrary \cite{Note3}.

Omitting unentangled qubits in $|0\ket$ states, the resulting boundary causal-cone state is
\begin{equation}
	|\tilde{\Psi}_\beta\ket = \frac{1}{2}\sum_{\nu,\mu=0,1}
	  e^{\mri\frac{\theta}{2}(-1)^{\nu+\mu}} |g_\nu\ket_\A\otimes|g_\mu\ket_\B,
\end{equation}
with $\theta=\arccos\beta$ and $|g_\nu\ket:=(\hat{V}|\nu\ket)^{\otimes T'}$. The reduced states for subsystems $\A$ and $\B$ are now unitarily equivalent to
Eq.~\eqref{eq:PauliLower-state} with $K=T'$.

Thus, no single $\B$-qubit exposes $\beta$ anymore. Moreover, $|\bra 0|\hat{V}^\dag\hs^\alpha\hat{V}|1\ket|=\sqrt{2/3}$ for every Pauli axis $\alpha\in\{x,y,z\}$ such that information about $\beta$ from any Pauli setting, including the all-$x$ setting, is exponentially suppressed with respect to $T=\Theta(\log\ell)$. This construction still results in the exponential lower bound 
$\Nshots^\text{Pauli}=\Omega\big((3/2)^{T/4}/\veps^2\big)$ even for nonuniform and adaptively chosen single-qubit Pauli axes while, still, $\Nshots^\text{replica}=\O(1/\veps^2)$.

\section{Clements-scheme implementation of the FFT}\label{appx:Uf}
\subsection{Setup and result}
The fermionic Fourier transform (FFT) $\hU_f$ as defined in Eq.~\eqref{eq:FFT} is the operator that realizes the single-particle basis transformation $u_f$ from Eq.~\eqref{eq:FFT-1particle} on the many-body Hilbert space of the $n$ considered qubits. Using Givens rotations $G_k:=G(\theta_k,\phi_k)$ \cite{Golub1996} with
\begin{equation}\label{eq:GivensRot}
	G(\theta,\phi)=\Pmatrix{\phantom{-e^{-\mri\phi}}\cos\theta & e^{i\phi}\sin\theta \\ -e^{-\mri\phi}\sin\theta & \phantom{e^{i\phi}}\cos\theta}
\end{equation}
according to the Clements scheme \cite{Clements2016-3}, we can sequentially nullify all off-diagonal elements of the $n\times n$ matrix $u_f$.

In the following, we will show that, as described briefly in Sec.~\ref{sec:shiftTest-implement}, this yields an implementation of $\hU_f^\dag$ with $n$ single-qubit $Z$ rotations
\begin{equation}\label{eq:Rz}
	\hR^z_j(\alpha_j):=e^{-\mri\frac{\alpha_j}{2}\hs^z_j}
\end{equation}
for the remaining diagonal matrix elements
and an $n$-layer brickwork circuit of $n(n-1)/2$ Givens-rotation two-qubit gates $\hG_k:=\hG_{j_k,j_k+1}(\theta_k, \phi_k)$ with
\begin{equation}\label{eq:Givens-mb-appx}\textstyle
	\hG_{i,j}(\theta, \phi) = \hR^{z\dag}_i(\phi+\frac{\pi}{2}) \, \hR^{xy}_{i,j}(\theta) \, \hR^z_i(\phi+\frac{\pi}{2})
\end{equation}
and the $XY$ interaction gates \cite{Makhlin1999-398,Imamoglu1999-83,Schuch2003-67,Steffen2006-313}
\begin{equation}\label{eq:Rxy}
	\hR^{xy}_{i,j}(\theta) = e^{-\mri \frac{\theta}{2} \left(\hs^x_i \hs^x_j + \hs^y_i \hs^y_j\right)}.
\end{equation}
The inverse FFT then takes the form
\begin{equation}\label{eq:FFT-brickwork}
	\hU_f^\dag=(\hG^\dag_2\hG^\dag_3\hG^\dag_7\dots)\Big(\prod_{j=1}^n \hR^z_j(\alpha_j)\Big)(\dots\hG_6\hG_5\hG_4\hG_1)
\end{equation}
as illustrated in Fig.~\ref{fig:FFT}.

\subsection{Clements decomposition of \texorpdfstring{$u_f$}{uf}}\label{appx:uf}
\begin{figure}[t]
	\includegraphics[width=\columnwidth]{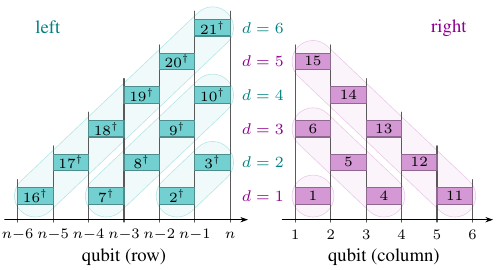}
	\caption{\label{fig:FFT-Givens}Diagonalization of the single-particle basis transformation matrix $u_f$ according to the Clements scheme is accomplished by products of Givens rotations that are to be applied to the left and right of $u_f$ as specified in Eq.~\eqref{eq:ufDiag}. The corresponding two-qubit gates \eqref{eq:Givens-mb-appx} on the many-body Hilbert space form a brickwork circuit as shown in Fig.~\ref{fig:FFT}b.}
\end{figure}

With row and column indices $r$ and $c$ for the single-particle transformation matrix $u_f$, we can label subdiagonals by indices
\begin{equation}
	d = c + n - r=1,2,\dotsc,n-1
\end{equation}
such that the $d=1$ subdiagonal consists of the bottom-left element $[u_f]_{n,1}$.
Following the Clements scheme \cite{Clements2016-3} and starting from $d=1$, we traverse all subdiagonal matrix elements as shown in Fig.~\ref{fig:FFT}a. For odd $d$, we visit matrix elements going up-left along subdiagonal $d$ and nullify each matrix element by a Givens rotation $G_k$, where the $2\times2$ matrix \eqref{eq:GivensRot} acts on columns $c$ and $c+1$ (from the right). For even $d$, we go down-right along the subdiagonal and nullify by acting with an adjoint Givens rotation $G_k^\dag$ on rows $r-1$ and $r$. When completing subdiagonal $d=n-1$, we have actually also nullified all superdiagonal matrix elements as every upper-triangular normal matrix is automatically diagonal \cite{Horn2013}, i.e.,
\begin{equation}\label{eq:ufDiag}
	(\dots G^\dag_3 G^\dag_2) u_f (G_1 G_4 G_5 G_6 \dots)=\diag(e^{-\mri\alpha_1},\dotsc,e^{-\mri\alpha_n}).
\end{equation}
See also the illustration in Fig.~\ref{fig:FFT-Givens}.

\subsection{Fermionic many-body form of Givens gates \texorpdfstring{$\hG_k$}{Gk}}
To establish the form \eqref{eq:Givens-mb} of the corresponding two-qubit gates $\hG_k$ on the many-body Hilbert space, first note that, in accordance with Eq.~\eqref{eq:FFT}, a Givens rotation $G=G(\theta,\phi)$, acting on columns $j$ and $j+1$, implements the transformation
\begin{equation}\label{eq:Gaa}
	\Pmatrix{\ha^\dag_j\\\ha^\dag_{j+1}}\mapsto
	\Pmatrix{\ta^\dag_j\\\ta^\dag_{j+1}} = G \Pmatrix{\ha^\dag_j\\\ha^\dag_{j+1}}.
\end{equation}
We claim that the unitary operator
\begin{subequations}\label{eq:G-manybodyFermion}
\begin{equation}
	\hG=e^{\hK}\ \ \text{with}\ \
	\hK=\theta\left(e^{\mri\phi}\ha^\dag_{j+1}\ha^\pdag_j-e^{-\mri\phi}\ha^\dag_j\ha^\pdag_{j+1}\right)
\end{equation}
is the corresponding transformation on the many-body Hilbert space such that
\begin{equation}
	\Pmatrix{\ta^\dag_j\\\ta^\dag_{j+1}} = \hG\Pmatrix{\ha^\dag_j\\\ha^\dag_{j+1}}\hG^\dag
	\equiv \Pmatrix{\hG\,\,\,\ha^\dag_j\,\,\hG^\dag \\\hG\ha^\dag_{j+1}\hG^\dag}
\end{equation}
\end{subequations}
and $\hG\ha^\dag_i\hG^\dag=\ha^\dag_i$ for all $i\neq j,j+1$.

For the proof, let $\hG(t):=e^{t\hK}$, which is unitary as $\hK=-\hK^\dag$ is skew-Hermitian, and let
\begin{equation}
	\Pmatrix{\ha^\dag_j(t)\\\ha^\dag_{j+1}(t)}:=\hG(t)\Pmatrix{\ha^\dag_j\\\ha^\dag_{j+1}}\hG^\dag(t).
\end{equation}

Evaluating the commutators $[\hK,\ha^\dag_j]$ and $[\hK,\ha^\dag_{j+1}]$, one finds
\begin{equation}\label{eq:dtat}
	\frac{\ud}{\ud t}\Pmatrix{\ha^\dag_j(t)\\\ha^\dag_{j+1}(t)}
	=\underbrace{\theta\Pmatrix{ & e^{i\phi} \\ -e^{-\mri\phi} & }}_{=:K}\Pmatrix{\ha^\dag_j(t)\\\ha^\dag_{j+1}(t)}
\end{equation}
such that, as $\hG=\hG(t=1)$,
\begin{align}\nonumber
	\hG&\Pmatrix{\ha^\dag_j\\\ha^\dag_{j+1}}\hG^\dag
	\stackrel{\eqref{eq:dtat}}{=}e^K\Pmatrix{\ha^\dag_j\\\ha^\dag_{j+1}}\\
	&=\Pmatrix{\phantom{-e^{-\mri\phi}}\cos\theta & e^{i\phi}\sin\theta \\ -e^{-\mri\phi}\sin\theta & \phantom{e^{i\phi}}\cos\theta}\Pmatrix{\ha^\dag_j\\\ha^\dag_{j+1}}.
\end{align}
The comparison of this result with Eqs.~\eqref{eq:GivensRot} and \eqref{eq:Gaa} confirms the claim \eqref{eq:G-manybodyFermion}.

\subsection{Pauli many-body form of Givens gates \texorpdfstring{$\hG_k$}{Gk}}
As we exclusively apply Givens rotations $G$ that either act on two adjacent rows or two adjacent columns of the single-particle basis transformation matrix in Appx.~\ref{appx:uf}, the corresponding Givens gates $\hG=e^{\hK}$ in Eq.~\eqref{eq:G-manybodyFermion} act on neighboring fermionic modes. Furthermore, the associated generator $\hK$ is bilinear in the fermionic creation and annihilation operators such that, when expressing them in terms of Pauli operators according to the Jordan-Wigner transformation \eqref{eq:JordanWigner}, no non-local Jordan-Wigner strings occur. Specifically,
\begin{align}\label{eq:K-manybody-spsm}
	\hK&\stackrel{\eqref{eq:JordanWigner},\eqref{eq:G-manybodyFermion}}{=}
	\theta\left(e^{\mri\phi}\hs^-_{j+1}\hs^+_j-e^{-\mri\phi}\hs^-_j\hs^+_{j+1}\right)\\ \nonumber
	&=\frac{i\theta}{2}\Big(
	       \ \ \sin\phi\left[\hs^x_j\hs^x_{j+1}+\hs^y_j\hs^y_{j+1}\right]\\
	&\qquad\ \ +\cos\phi\left[\hs^y_j\hs^x_{j+1}-\hs^x_j\hs^y_{j+1}\right]\Big)
	\label{eq:K-manybody-sxsy}
\end{align}

Finally, we want to express $\hG=e^{\hK}$ in terms of single-qubit $Z$ rotations \eqref{eq:Rz} and two-qubit $XY$ interaction gates \eqref{eq:Rxy}. To establish the claimed relation \eqref{eq:Givens-mb-appx}, note that
\begin{align}
	&\hR_j^z(\gamma)\hs_j^+\hR_j^{z\dag}(\gamma)=e^{-\mri\gamma}\hs_j^+\quad\text{such that}\\
	&\hR_j^z(\gamma)\hK\hR_j^{z\dag}(\gamma)\stackrel{\eqref{eq:K-manybody-spsm}}{=}
	 \theta\left(e^{\mri(\phi-\gamma)}\hs^-_{j+1}\hs^+_j-H.c.\right).
\end{align}
Choosing $\gamma=\phi+\frac{\pi}{2}$ achieves $\sin(\phi-\gamma)=-1$ and $\cos(\phi-\gamma)=0$, i.e.,
\begin{equation}\textstyle
	\hR_j^z(\phi+\frac{\pi}{2})\,\hG\,\hR_j^{z\dag}(\phi+\frac{\pi}{2})\stackrel{\eqref{eq:K-manybody-sxsy}}{=}
	e^{-\frac{i\theta}{2}\left(\hs^x_j\hs^x_{j+1}+\hs^y_j\hs^y_{j+1}\right)},
\end{equation}
which proves Eq.~\eqref{eq:Givens-mb-appx}.

\subsection{FFT circuit}
With the Clements decomposition \eqref{eq:ufDiag} of $u_f$,
\begin{equation*}
	 u_f^\dag=(G_1 G_4 G_5 G_6 \dots)\diag(e^{\mri\alpha_1},\dotsc,e^{\mri\alpha_n})(\dots G^\dag_3 G^\dag_2).
\end{equation*}
We can now use the established relation of the Givens rotations $G_k$ to the two-qubit Givens gates $\hG_k$ [Eqs.~\eqref{eq:Givens-mb} and \eqref{eq:Givens-mb-appx}], and $Z$ rotations $\hR_j^z(\alpha_j)$ [Eq.~\eqref{eq:Rz}] to implement the remaining phase factors on the diagonal of the single-particle basis transformation matrix with $\hR_j^z(\alpha_j)\ha_j^\dag\hR_j^{z\dag}(\alpha_j)=e^{\mri\alpha_j}\ha_j^\dag$. Finally, note that, when implementing a product of two single particle transformations in the many-body Hilbert space, the order of transformations is reversed in the sense that
\begin{equation}
	\hG_b\hG_a \ha_q^\dag \hG_a^\dag\hG_b^\dag \stackrel{\eqref{eq:FFT}}{=}  \sum_j [G_a G_b]_{q,j}\ha_j^\dag.
\end{equation}
With this, we arrive at the $n$-layer brickwork circuit \eqref{eq:FFT-brickwork} for the inverse FFT $\hU_f^\dag$ as illustrated in Fig.~\ref{fig:FFT} for $n=6$ qubits.

\section{Shift and flux operators}\label{appx:F}
This appendix proves the relation \eqref{eq:CFCfF},
\begin{equation*}
	\hC=e^{\mri\frac{\pi}{n}\hN\delta_{\hN,\text{even}}} \hF^\dag \hC_f \hF,
\end{equation*}
between the shift operator $\hC$ on $n$ qubits [Eq.~\eqref{eq:C}] and the fermionic shift operator $\hC_f$ [Eq.~\eqref{eq:Cf}] as mediated by the parity-controlled flux operator $\hF$ [Eq.~\eqref{eq:F}]. All three operators preserve the total particle number $N=\sum_{j=1}^n\nu_j$.

In the odd-$N$ sector, we have $\hF=\id$ and
\begin{align}\nonumber
	&\hC_f|\nu_1,\nu_2,\dotsc,\nu_n\ket
	\stackrel{\eqref{eq:ShiftSign}}{=}(-1)^{\nu_n\,(N-1)}|\nu_n,\nu_1,\dotsc,\nu_{n-1}\ket\\
	&=|\nu_n,\nu_1,\dotsc,\nu_{n-1}\ket=\hC|\nu_1,\nu_2,\dotsc,\nu_n\ket
\end{align}
in agreement with Eq.~\eqref{eq:CFCfF}.

In the even-$N$ sector, we have 
\begin{align}\nonumber
	&\hC_f|\nu_1,\nu_2,\dotsc,\nu_n\ket
	\stackrel{\eqref{eq:ShiftSign}}{=}(-1)^{\nu_n}|\nu_n,\nu_1,\dotsc,\nu_{n-1}\ket\\
	&=(-1)^{\nu_n}\hC|\nu_1,\nu_2,\dotsc,\nu_n\ket
\end{align}
such that
\begin{align*}
	&\hF^\dag \hC_f\hF|\nu_1,\nu_2,\dotsc,\nu_n\ket\\
	&\stackrel{\eqref{eq:F}}{=}
	(-1)^{\nu_n}e^{\mri\frac{\pi}{n}\sum_j j\nu_j}\hF^\dag|\nu_n,\nu_1,\dotsc,\nu_{n-1}\ket\\
	&\stackrel{\eqref{eq:F}}{=}
	(-1)^{\nu_n}e^{\mri\frac{\pi}{n}\left[\nu_n(n-1) + \sum_{j<n} (j-(j+1))\nu_j\right]}\\
	&\qquad\times |\nu_n,\nu_1,\dotsc,\nu_{n-1}\ket\\
	&\,\,=e^{-\mri\frac{\pi}{n}N} \hC|\nu_1,\nu_2,\dotsc,\nu_n\ket
\end{align*}
in agreement with Eq.~\eqref{eq:CFCfF}.

\section{Quantum circuit for the parity-controlled flux operator}\label{appx:F-implement}
The parity-controlled flux operator $\hF$ in Eq.~\eqref{eq:F} is diagonal in the computational basis and can be realized as a circuit of $\Theta(n)$ finite-range CNOT gates and single-qubit $Z$ rotations as discussed in the following.

Using Pauli operators, we can write its generator in the form
\begin{align}\nonumber
	&\frac{\pi}{n}\delta_{\hN,\text{even}}\sum_{j=1}^n j\hn_j
	 =\frac{\pi}{4n}\biggl(\id + \prod_{j=1}^n \hs^z_j\biggr)\sum_{j=1}^n j(\id-\hs^z_j)\\
	&=\frac{\pi}{4n}\biggl(S -\!\sum_j j\hs^z_j + S\prod_j\hs^z_j -\! \sum_j j \prod_{i\neq j}^n \hs^z_i \biggr)
\end{align}
with $S=\sum_{j=1}^n j = {n(n+1)}/{2}$ such that $\hF$ can be written as a sequence of $2n+1$ commuting unitary rotations
\begin{equation}\label{eq:Frotations}
	\hF=e^{\mri \phi} \biggl( \prod_{j=1}^n e^{-\mri\frac{\theta_j}{2}\hs^z_j} \biggr) e^{\mri\phi\hP}\Biggl( \prod_{j=1}^n e^{-\mri\frac{\theta_j}{2}\hP_{\setminus j}} \Biggr)
\end{equation}
with rotation angles $\theta_j:=\frac{\pi j}{2n}$, $\phi=\frac{\pi(n+1)}{8}$, and the parity operators
\begin{equation}
	\hP:=\prod_{j=1}^n \hs^z_j\quad\text{and}\quad
	\hP_{\setminus j}:=\prod_{i\neq j}\hs^z_i=\hs^z_j \hP.
\end{equation}

The first factor in Eq.~\eqref{eq:Frotations} is implemented by applying $n$ single-qubit $Z$ rotations $\hR^z_j(\theta_j)$.

For the remaining two factors, we need to access the global parity $\hP$ and all-but-one parities $\hP_{\setminus j}$. We can implement these by traversing the $n$ sites three times, applying nearest and next-nearest neighbor CNOT gates and single-qubit $Z$ gates:
\begin{enumerate}
 \item 
 Applying a sequence of $n-1$ nearest-neighbor CNOT gates from right to left to a basis state,
 \begin{gather}\nonumber
 	\CNOT_{2,1}\CNOT_{3,2}\dotsb\CNOT_{n,n-1}|\nu_1,\nu_2,\dotsc,\nu_n\ket\\
 	=|P_1,P_2,P_3,\dotsc,P_n\ket,
 \end{gather}
 each qubit $j$ now holds the \emph{right-partial parity}
 \begin{equation}
	P_j:=\nu_j\oplus\nu_{j+1}\oplus\dotsb\oplus\nu_{n}\in\{0,1\},
 \end{equation}
 where $\oplus$ denotes addition modulo two and $P_1$ is the global parity.
 \item
 Applying the $Z$ rotation $\hR^z_1(\theta)$ with $\theta=-\pi(n+1)/4$ on site 1 implements the second factor in Eq.~\eqref{eq:Frotations}.
 \item
 Applying the $Z$ rotation $\hR^z_2(\theta_1)$  on site 2 implements the term $\exp(-\mri\frac{\theta_1}{2}\hP_{\setminus 1})$ from the third factor in Eq.~\eqref{eq:Frotations} as $\hP_2=\hP_{\setminus 1}$.
 \item
 Applying $\CNOT_{2,1}$ brings us back to the state $|\nu_1,P_2,P_3,\dotsc,P_n\ket$.
 \item
 Moving to the right over sites $j=2,3,\dotsc,n-1$, we can implement the terms $\exp(-\mri\frac{\theta_j}{2}\hP_{\setminus j})$ from Eq.~\eqref{eq:Frotations}. Specifically, we start from the state
 \begin{equation}
	|P'_1,\dotsc,P'_{j-1},P_j,P_{j+1},\dotsc,P_n\ket,
 \end{equation}
 where
 \begin{equation}
	P'_j:=\nu_1\oplus\nu_2\oplus\dotsb\oplus\nu_j\in\{0,1\}
 \end{equation}
 is the \emph{left-partial parity} for site $j$.
 Acting with $\CNOT_{j-1,j+1}$ takes us to the state $|\dotsc,P'_{j-1},P_j,P'_{j-1}\oplus P_{j+1},\dotsc\ket$
 such that application of $\hR^z_{j+1}(\theta_j)$ implements the term $\exp(-\mri\frac{\theta_j}{2}\hP_{\setminus j})$ from Eq.~\eqref{eq:Frotations}. The gates $\CNOT_{j-1,j+1}\CNOT_{j+1,j}$ then take us to the next state $|\dotsc,P'_{j-1},P'_j,P_{j+1},\dotsc\ket$.
 \item
 When reaching the right end of the chain, the system state is $|P'_1,\dotsc,P'_{n-1},\nu_n\ket$ such that application of $\hR^z_{n-1}(\theta_n)$ implements the final term $\exp(-\mri\frac{\theta_n}{2}\hP_{\setminus n})$ as $\hP'_{n-1}=\hP_{\setminus n}$.
 \item
 A final clean-up sweep to the left with $n-2$ CNOTs,
 \begin{gather}\nonumber
 	\CNOT_{1,2}\dotsb\CNOT_{n-2,n-1}|P'_1,\dotsc,P'_{n-2},P'_{n-1},\nu_n\ket\\
 	=|\nu_1,\nu_2,\dotsc,\nu_{n-1},\nu_n\ket,
 \end{gather}
 restores the initial state.
\end{enumerate}

The circuit comprises $5n-8$ CNOT gates and $2n+1$ single-qubit $Z$ rotations.
For $n=5$, this quantum circuit is shown in Fig.~\ref{fig:C}.

\section{Custom circuits for small-\texorpdfstring{$n$}{n} shift-operator measurements}\label{appx:small-n}
In Secs.~\ref{sec:shiftTest-FFT} and \ref{sec:shiftTest-implement}, we have described how a measurement of the cyclic shift operator $\hC$ [Eq.~\eqref{eq:C}] can be implemented for any number $n$ of qubits (qubit copies) by transforming to its eigenbasis through application of the parity-controlled flux operator $\hF$ and the inverse FFT $\hU_f^\dag$ according to Eq.~\eqref{eq:FUDUF}, and then measuring in the computational basis. The implementation requires 
a total of $n(n+9)/2-8$ CNOT and $XY$ gates (on superconducting devices when ignoring SWAP gates) or a total of $n(n+4)-8$ CNOT or MS gates (on trapped-ion devices).

For specific (small) $n$, one can generally find alternative circuits that diagonalize the shift operator $\hC$ with fewer gates. One route is to break conservation of the fermionic particle number (magnetization) $N=\sum_{j=1}^n\nu_j$, which the flux operator $\hF$ and FFT $\hU_f$ of the general-$n$ scheme obey. Adaptive schemes using mid-circuit measurements would allow for further gate count reductions, and could be pursued in future work.

\subsection{Case \texorpdfstring{$n=2$}{n=2}}\label{appx:small-2}
As already pointed out in Sec.~\ref{sec:shiftTest-small-n} and shown in Fig.~\ref{fig:shiftTest-small-n}a, the shift operator $\hC=\SWAP$ for $n=2$ is diagonalized by $\Had_1\, \CNOT_{1,2}$ [Eq.~\eqref{eq:Cdiag-n2}] as
\begin{align*}
	&\CNOT_{1,2}\,\Had_1\quad\Omega\quad\Had_1\, \CNOT_{1,2}\\
	=\, &\CNOT_{1,2}\,\Had_1\, \text{CZ}_{2,1}\, \Had_1\, \CNOT_{1,2}\\
	=\, &\CNOT_{1,2}\ \ \ \,\CNOT_{2,1}\ \ \ \CNOT_{1,2} = \SWAP.
\end{align*}

This scheme uses only one instead of the three CNOT+$XY$ gates or four CNOTs in the general-$n$ scheme [Eqs.~\eqref{eq:cost-SC} and \eqref{eq:cost-ionTrap}]. 
Obviously, $\Had_1$ breaks the particle-number conservation, in this way, exploiting the three-fold degeneracy of the $\hC$ eigenvalue $1$ to reduce the gate count.

\subsection{Case \texorpdfstring{$n=3$}{n=3}}\label{appx:small-3}
The cyclic shift operator $\hC$ for $n=3$ qubits leaves the computational basis states $|0,0,0\ket$ and $|1,1,1\ket$ invariant. In the remaining two Hamming-weight subspaces it implements the 3-cycles
\begin{gather*}
	|0,1,1\ket \to |1,0,1\ket \to |1,1,0\ket \to |0,1,1\ket\,\ \text{for}\,\ P=0,\\
	|1,0,0\ket \to |0,1,0\ket \to |0,0,1\ket \to |1,0,0\ket\,\ \text{for}\,\ P=1
\end{gather*}
in terms of the computational basis states $|\nu_1,\nu_2,\nu_3\ket$, where the parity is $P=\nu_1\oplus\nu_2\oplus\nu_3$.

To efficiently diagonalize $\hC$, we can isolate the (invariant) parity $P$ in qubit 3 and decouple the two cycles from $P$ using three CNOT gates:
\begin{equation*}
	\CNOT_{1,3}\CNOT_{3,2}\CNOT_{2,1}|\nu_1,\nu_2,\nu_3\ket
	\!=\!|\nu_1\oplus\nu_2,\nu_2\oplus\nu_3,P\ket
\end{equation*}
such that, independent of $P\in\{0,1\}$, the operator
\begin{equation}
	\hC':=\hU \hC \hU^\dag\ \ \text{with}\ \
	\hU=\CNOT_{1,3}\CNOT_{3,2}\CNOT_{2,1}
\end{equation}
now causes the \emph{same} 3-cycle on qubits 1 and 2:
\begin{equation}
	|1,0\ket \to |1,1\ket \to |0,1\ket \to |1,0\ket,
\end{equation}
leaving the state $|0,0\ket$ and qubit 3 invariant.

In other words,
\begin{gather*}
	\hC'=\hat{c}\otimes\id\quad\text{with}\\
	\hat{c}\!=\!|0,0\ket\bra 0,0|+|1,1\ket\bra 1,0|+|0,1\ket\bra 1,1|+|1,0\ket\bra 0,1|,
\end{gather*}
and the two-qubit operator $\hat{c}$ can be diagonalized by the unitary
\begin{equation}
	\hat{V} = \Pmatrix{
	1 & 0 & 0 & 0 \\
	0 & \frac{1}{\sqrt{3}} & \frac{1}{\sqrt{3}} & \frac{1}{\sqrt{3}} \\
	0 & \frac{1}{\sqrt{3}} & \frac{\zeta^*}{\sqrt{3}} & \frac{\zeta}{\sqrt{3}} \\
	0 & \frac{1}{\sqrt{3}} & \frac{\zeta}{\sqrt{3}} & \frac{\zeta^*}{\sqrt{3}} }
	\quad\text{with}\quad
	\zeta = e^{\mri 2\pi/3},
\end{equation}
where we have used the basis $(|0,0\ket,|1,0\ket,|0,1\ket,|1,1\ket)$.
We then have,
\begin{gather}
	\hC = \hU^\dag (\hat{c}\otimes\id) \hU
	= \hU^\dag\big((\hat{V}^\dag\Omega\hat{V})\otimes\id\big) \hU\\
	\text{with}\quad
	\Omega=\diag(1,1,\zeta,\zeta^*).
\end{gather}

A standard circuit for the implementation of a general two-qubit unitary like $\hat{V}$ comprises three CNOT gates, one $Z$ and two $Y$ single-qubit rotations, as well as four general single-qubit gates \cite{Vatan2004-69,Shende2004-69}. Thus, applying the circuit for $(\hat{V}\otimes\id) \hU$ as shown in Fig.~\ref{fig:shiftTest-small-n}b, we can measure the shift operator using $3+3=6$ CNOT gates instead of the 10 CNOT+$XY$ gates or 13 CNOTs in the general-$n$ scheme [Eqs.~\eqref{eq:cost-SC} and \eqref{eq:cost-ionTrap}].

Explicitly, $\hat{V}$ is realized by the circuit
\begin{align}\nonumber
	\hat{V}=&e^{\mri 3\pi/8}(\hU_1'\otimes \hU_2')\,\CNOT_{1,2}\Big(\hR_y(\theta_y')\otimes\id\Big)\CNOT_{2,1}\\
	        &\times\Big(\hR_y(\theta_y)\otimes\hR_z(\theta_z)\Big)\,\CNOT_{1,2}(\hU_1\otimes \hU_2)
\end{align}
with the three internal rotation angles
\begin{subequations}
\begin{align}
	\theta_y&=+2.67448671,\quad
	\theta_y'=-0.91670787,\\
	\theta_z&=-1.12201442.
\end{align}
\end{subequations}
The four single-qubit gates $\hU_1,\hU_2,\hU_1'$, and $\hU_2'$ can be implemented as
\begin{equation}
	\hU(\alpha,\beta,\gamma)=\hR_z(\alpha)\hR_y(\beta)\hR_z(\gamma)
\end{equation}
with the following rotation angles.

\begin{center}
\setlength{\tabcolsep}{1ex}
\renewcommand{\arraystretch}{1.5}
\begin{tabular}{cccc}
Gate     &   $\alpha$ &    $\beta$ &   $\gamma$\\\hline
$\hU_1$  &+4.05175685 &-3.79914538 &-0.61709003\\
$\hU_2$  &-5.62255317 &-0.65755273 &-4.80588024\\
$\hU_1'$ &-0.61709003 &+3.79914538 &-0.66063213\\
$\hU_2'$ &-1.66428759 &-0.65755273 &-0.91016419
\end{tabular}
\end{center}

\subsection{Case \texorpdfstring{$n=4$}{n=4}}\label{appx:small-4}
In the general-$n$ scheme of Sec.~\ref{sec:shiftTest-all-n}, we applied the parity-controlled flux operator $\hF$ to correct for the fermionic sign in Eq.~\eqref{eq:ShiftSign} that distinguishes the fermionic shift operator $\hC_f$ from the actual qubit shift operator $\hC$. For $n=4$, two-qubit gate counts can be reduced by noting that it suffices to find a diagonal operator $\hD$ such that $\hD\hC\hD^\dag$ differs from the fermionic shift operator $\hC_f$ by a phase that only depends on the conserved particle number (Hamming weight) $\hN=\sum_{j=1}^n\hn_j$.

An efficient choice for $n=4$ is
\begin{align}\label{eq:n4-D}
	&\hD
	=e^{\mri\pi\left(\frac{3}{8}\hs_1^z-\frac{1}{4}\hs_2^z-\frac{3}{8}\hs_3^z
	                 +\frac{1}{8}\hs_1^z\hs_2^z-\frac{3}{8}\hs_1^z\hs_3^z +\frac{3}{8}\hs_3^z\hs_4^z\right)}\\
	&\textstyle\nonumber
	 =\hR_1^z(\frac{-3\pi}{4})\hR_2^z(\frac{\pi}{2})\hR_3^z(\frac{3\pi}{4})
	  \hR_{1,2}^{zz}(\frac{-\pi}{4})\hR_{1,3}^{zz}(\frac{3\pi}{4})\hR_{3,4}^{zz}(\frac{-3\pi}{4})
\end{align}
with the $ZZ$ interaction gates
\begin{equation}\label{eq:Rzz}
	\hR^{zz}_{i,j}(\theta) = e^{-\mri \frac{\theta}{2} \hs_i^z\hs_j^z}
\end{equation}
and the single-qubit rotations $\hR_j^z$ as defined in Eq.~\eqref{eq:Rz}.

Computational basis states \eqref{eq:FockStates} are $\hD$ eigenstates, $\hD|\nu_1,\nu_2,\nu_3,\nu_4\ket=D_\vn|\nu_1,\nu_2,\nu_3,\nu_4\ket$, where the eigenvalue ratio of  states related by a cyclic shift \eqref{eq:C} is
\begin{subequations}
\begin{gather}
	\frac{D_{\nu_4,\nu_1,\nu_2,\nu_3}}{D_{\nu_1,\nu_2,\nu_3,\nu_4}} = h_N\,(-1)^{\nu_4(N-1)} \quad\text{with}\\
	(h_0,h_1,h_2,h_3,h_4)=(1,\mri,\mri,1,-1).
\end{gather}
\end{subequations}
Comparing this to Eq.~\eqref{eq:ShiftSign} which, for $n=4$, reads
$\hC_f|\nu_1,\nu_2,\nu_3,\nu_4\ket = (-1)^{\nu_4(N-1)}|\nu_4,\nu_1,\nu_2,\nu_3\ket$, we find
\begin{equation}\label{eq:n4-DCD}
	\hD\hC\hD^\dag=h_{\hN}\hC_f.
\end{equation}

Hence, using the expression \eqref{eq:Cf} for $\hC_f$ and that the FFT \eqref{eq:FFT} conserves $\hN$, we see that the 4-qubit shift operator $\hC$ is diagonalized by the measurement unitary $\hU_f^\dag\hD$ such that
\begin{equation}\label{eq:n4-U4}
	\hC = \hD^\dag \hU_f\cdot h_{\hN}\, e^{-\mri\frac{\pi}{2}(\hn_1+2\hn_2+3\hn_3+4\hn_4)}\cdot \hU_f^\dag \hD.
\end{equation}
Applying the operator $\hD$, then applying the inverse FFT $\hU_f^\dag$, and finally measuring the $n=4$ qubits in the computational basis to obtain the occupation numbers $\nu_j\in\{0,1\}$, the associated $\hC$ eigenvalue is
\begin{equation}\label{eq:n4-omega-k}
	\omega=h_N\, e^{-\mri\frac{\pi}{2}(\nu_1+2\nu_2+3\nu_3+4\nu_4)}
	\ \ \text{with}\ \  N=\sum_j\nu_j
\end{equation}
in analogy to Eq.~\eqref{eq:omega-k}.

The inverse FFT $\hU_f^\dag$ can again be implemented according to the Clements scheme \eqref{eq:FFT-brickwork}. Specifically, it takes the form
\begin{equation}
	\hU_f^\dag	=(\hG_2^\dag\hG_3^\dag)	\Big(\prod_{j=1}^4 \hR_j^z(\alpha_j)\Big)(\hG_6\hG_5\hG_4\hG_1)
\end{equation}
with $(\alpha_1,\alpha_2,\alpha_3,\alpha_4)=\left(-\frac{\pi}{4},-\frac{\pi}{2},-\pi,\frac{\pi}{4}\right)$ and the Givens-rotation two-qubit gates
\begin{alignat*}{3}
	\hG_1&\textstyle=\hG_{1,2}(\frac{\pi}{4},0),\ 					&&\textstyle\hG_2=\hG_{2,3}(\arctan\frac{1}{\sqrt{2}},\frac{\pi}{4}),\\
	\hG_3&\textstyle=\hG_{3,4}(\frac{\pi}{3},\frac{\pi}{4}),\		&&\textstyle\hG_4=\hG_{3,4}(\frac{\pi}{4},-\frac{\pi}{2}),\\
	\hG_5&\textstyle=\hG_{2,3}(\arctan\sqrt{2},-\frac{\pi}{4}),\ \	&&\textstyle\hG_6=\hG_{1,2}(\frac{\pi}{3},-\frac{\pi}{2})
\end{alignat*}
according to the definition \eqref{eq:Givens-mb-appx}.

The full circuit is shown in Fig.~\ref{fig:shiftTest-small-n}c. The FFT for $n=4$ comprises six Givens rotations, i.e., six $XY$ interactions, and the diagonal phase operator $\hD$ comprises three $ZZ$ rotations [Eq.~\eqref{eq:n4-D}]. On a superconducting device with a native $XY$ gates and tunable-angle $ZZ$ (or controlled-phase) entanglers, this gives a total of $6\,XY+3\,ZZ=9$ native two-qubit gates. If $ZZ$ rotations need to be implemented using two CNOT or fixed-angle CZ entanglers, one gets a total of $6\,XY+6\,\text{CNOT/CZ}=12$ native two-qubit gates. This is a reduction compared to the $18$ native two-qubit gates in the general-$n$ scheme according to Eq.~\eqref{eq:cost-SC}.

For trapped-ion computers, the native two-qubit entanglers are M\o{}lmer-S\o{}rensen (MS) gates on systems like IonQ Aria or $ZZ$ interaction gates on systems like IonQ Forte. In either case, each of the six Givens rotations can be implemented using two native entangling gates and single-qubit rotations.
With tunable interaction angles, each of the three $ZZ$ rotations in $\hD$ requires only one native entangling gate, giving a total of $6\times2+3=15$ native
two-qubit gates instead of the 24 gates in the general-$n$ scheme according to Eq.~\eqref{eq:cost-ionTrap}.

These counts again omit potential routing overhead. In particular, the $R^{zz}_{1,3}$ interaction is nonadjacent on a linear nearest-neighbor layout; any required SWAP or routing gates are not included here.

The achieved reduction for $n=4$ exploits the sparse structure of the diagonal phase operator $\hD$ in Eq.~\eqref{eq:n4-D}. A straightforward extension to arbitrary $n$ involves a dense set of pairwise $ZZ$ phase rotations, i.e., it has $\Theta(n^2)$ two-qubit gates and, hence, no asymptotic advantage compared to the parity-controlled flux operator $\hF$ from Secs.~\ref{sec:shiftTest-FFT} and \ref{sec:shiftTest-implement}, which requires only $5n-8=\Theta(n)$ CNOT gates.

\end{document}